\documentclass[twocolumn,english]{revtex4-2}
\usepackage{lmodern}
\usepackage{lmodern}

\usepackage[T1]{fontenc}
\usepackage[latin9]{inputenc}
\usepackage{babel}
\usepackage{float}
\usepackage{amsmath}
\usepackage{amssymb}
\usepackage{graphicx}
\usepackage[unicode=true,pdfusetitle,
 bookmarks=true,bookmarksnumbered=false,bookmarksopen=false,
 breaklinks=false,pdfborder={0 0 1},backref=false,colorlinks=false]
 {hyperref}

\makeatletter

\providecommand{\tabularnewline}{\\}

\usepackage{babel}

\makeatother

\begin{document}
\title{Analytic solutions of the nonlinear radiation diffusion equation with
an instantaneous point source in non-homogeneous media}
\author{Menahem Krief}
\email{menahem.krief@mail.huji.ac.il}

\address{Racah Institute of Physics, The Hebrew University, 9190401 Jerusalem,
Israel}
\begin{abstract}
Analytic solutions to the nonlinear radiation diffusion equation with
an instantaneous point source for a non-homogeneous medium with a
power law spatial density profile, are presented. The solutions are
a generalization of the well known solutions for a homogeneous medium.
It is shown that the solutions take various qualitatively different
forms according to the value of the spatial exponent. These different
forms are studied in detail for linear and non linear heat conduction.
In addition, by inspecting the generalized solutions, we show that
there exist values of the spatial exponent such the conduction front
has constant speed or even accelerates. Finally, the various solution
forms are compared in detail to numerical simulations, and a good
agreement is achieved.
\end{abstract}
\maketitle

\section{Introduction}

Radiative heat waves are an important phenomena in many astrophysical
and laboratory high energy density plasmas \cite{lindl2004physics,back2000diffusive,robey2001experimental,bailey2015higher,falize2011similarity,hurricane2014fuel,cohen2020key,heizler2021radiation}.
As a result, analytic solutions for the radiative heat equation play
a key role in the analysis and design of high energy density experiments
\cite{sigel1988x,back2000diffusive,keiter2008radiation,lindl1995development,heizler2021radiation}
and in the process of verification and validation of computer simulations
\cite{reinicke1991point,shestakov1999time,calder2002validating,krumholz2007equations,gittings2008rage,coggeshall1986lie,coggeshall1991analytic,lowrie2007radiative,ramsey2018converging,ramsey2019piston,bingjing1996benchmark,ruby2019boundary,modelevsky2021revisiting}. 

Analytic solutions for radiative heat waves with externally applied
boundary conditions were developed in the seminal work of Marshak
\cite{marshak1958effect} which was further generalized in Refs. \cite{pakula1985self,kaiser1989x,hammer2003consistent,saillard2010principles,lane2013new,shussman2015full,heizler2016self,cohen2018modeling}.
Analytical solutions to the nonlinear diffusion equation with an instantaneous
point source for a homogeneous medium was developed in the seminal
works of Zel'dovich et al. \cite{zel1959propagation,zeldovich1967physics,barenblatt1996scaling}
and Pattle \cite{pattle1959diffusion}. A solution for the linear
diffusion equation in a non-homogeneous medium was developed in Ref.
\cite{o1985analytical}, in order to describe diffusion on fractal
objects.

In this work we extend the solution of Zel'dovich et al. \cite{zel1959propagation,zeldovich1967physics,barenblatt1996scaling}
and Pattle \cite{pattle1959diffusion}, and develop analytic solutions
to the nonlinear radiation diffusion equation with an instantaneous
point source for a non-homogeneous medium with a power law spatial
density profile of the form $\rho_{0}r^{-\omega}$. Such
profiles are widely used, for example, in modeling the interior and
atmospheres of stars \cite{meszaros2002theories,tan2001trans,sapir2011non,katz2012non,sapir2013non}
and galaxies \cite{evans1994power,koopmans2009structure,schneider2013mass}.
These solutions are analyzed in detail for both linear and non linear
conduction. Different solution forms are examined for various ranges
of the spatial exponent $\omega$. Finally, the solutions are compared
with numerous numerical simulations.

\section{Statement of the problem}

In situations where radiation heat conduction dominates and hydrodynamic
motion is negligible, the material density is constant in time, and
the heat flow is supersonic. A comparison between the dynamics of
radiation conduction and compressible flow in the context of an instantaneous
point source, also known as the strong explosion problem \cite{sedov1946propagation,taylor1950formation,reinicke1991point,sedov1993similarity,waxman1993second,kamm2007efficient},
will be detailed below in Sec. \ref{subsec:Relation-to-the}. Assuming
a constant spatial density, the radiation diffusion equation in one
dimensional symmetry is given by:

\begin{equation}
\frac{\partial u}{\partial t}=-\frac{1}{r^{d-1}}\frac{\partial}{\partial r}\left(r^{d-1}F\right),\label{eq:main_eq}
\end{equation}
where $u\left(r,t\right)$ is the total energy per unit volume and
$d=1,2,3$ for planar, cylindrical and spherical symmetries, respectively.
The radiation energy flux obeys a Fick law:
\begin{equation}
F=-D\frac{\partial}{\partial r}\left(aT^{4}\right),\label{eq:fick}
\end{equation}
with $T$ the material temperature, $a$ the radiation constant, and
the radiation diffusion coefficient:

\begin{equation}
D=\frac{c}{3\kappa_{R}\rho},\label{eq:diffusion_coeff}
\end{equation}
where $c$ is the speed of light, $\kappa_{R}$ is the Rosseland mean
opacity and $\rho$ is the material mass density. The flux $F$ is,
in general, a non-linear function of $u$ and its derivative.

In this work, we assume power law opacity and energy equation of state,
in the common form \cite{hammer2003consistent,garnier2006self,shussman2015full,heizler2016self,heizler2021radiation}:

\begin{equation}
\frac{1}{\kappa_{R}\left(T,\rho\right)}=gT^{\alpha}\rho^{-\lambda},\label{eq:ross_opac_powerlaw}
\end{equation}

\begin{equation}
u\left(T,\rho\right)=fT^{\beta}\rho^{1-\mu}.\label{eq:eos}
\end{equation}
and an inhomogeneous density profile in a spatial power law form:
\begin{equation}
\rho\left(r\right)=\rho_{0}r^{-\omega}.\label{eq:rho_powerlaw}
\end{equation}
We note that in order for this density profile to contain a finite
total mass, one must have $\omega<d$. In addition, it is evident
from equations \eqref{eq:diffusion_coeff}-\eqref{eq:rho_powerlaw}
that the above power laws are equivalent to a single temperature and
spatial power law for the diffusion coefficient:
\begin{equation}
D\left(r,T\right)=D_{0}r^{-\omega'}T^{\alpha}.\label{eq:diffpow}
\end{equation}
with $D_{0}=\frac{gc}{3\rho_{0}^{\lambda+1}}$, $\omega'=\omega\left(\lambda+1\right)$.
The radiation diffusion coefficient as given in Eq. \eqref{eq:diffusion_coeff},
is a non-linear function of $u$, in terms of $T$. 

We consider an instantaneous point source, so that the initial energy
density profile is given by:

\begin{equation}
u\left(r,t=0\right)=Q\delta\left(\boldsymbol{r}\right),\label{eq:point_source}
\end{equation}
where $Q$ is the total initial energy, which is of course, constant
in time for an infinite system:
\begin{equation}
\int_{0}^{\infty}u\left(r,t\right)\mathcal{A}_{d}r^{d-1}dr=Q,\label{eq:conservation}
\end{equation}
where the areal coefficient is:
\begin{equation}
\mathcal{A}_{d}=\begin{cases}
1 & d=1\\
2\pi & d=2\\
4\pi & d=3
\end{cases}
\end{equation}

Using equations \eqref{eq:fick}-\eqref{eq:rho_powerlaw}, the diffusion
equation \eqref{eq:main_eq} takes the form:

\begin{equation}
\frac{\partial u}{\partial t}=\frac{1}{r^{d-1}}\frac{\partial}{\partial r}\left(Ar^{k+d-1}\left(r^{m}u\right)^{n}\frac{\partial}{\partial r}\left(r^{m}u\right)\right),\label{eq:dudt}
\end{equation}
where:
\begin{equation}
n=\frac{4+\alpha-\beta}{\beta},
\end{equation}

\begin{equation}
k=\omega\left(1+\lambda\right),
\end{equation}
\begin{equation}
m=\omega\left(1-\mu\right),
\end{equation}
\begin{equation}
A=\frac{16\sigma g}{3\beta f^{\frac{4+\alpha}{\beta}}\rho_{0}^{\lambda+1+\frac{\left(1-\mu\right)\left(\alpha+4\right)}{\beta}}}.
\end{equation}
By defining the auxiliary variable: 
\begin{equation}
w\left(r,t\right)=r^{m}u\left(r,t\right),
\end{equation}
 Eq. \eqref{eq:dudt} takes the simple from:

\begin{equation}
\frac{\partial w}{\partial t}=Ar^{m-d+1}\frac{\partial}{\partial r}\left(r^{k+d-1}w^{n}\frac{\partial w}{\partial r}\right),\label{eq:diff_eq_1d}
\end{equation}
It is seen from Eq. \eqref{eq:dudt} that the energy flux can be written
in terms of $w$ as:
\begin{equation}
F\left(r,t\right)=-Ar^{k}w^{n}\frac{\partial w}{\partial r}.\label{eq:flux_w}
\end{equation}
The energy conservation law in Eq. \eqref{eq:conservation}, written
in terms of $w$, reads:
\begin{equation}
\int_{0}^{\infty}w\left(r,t\right)\mathcal{A}_{d}r^{d-m-1}dr=Q.\label{eq:cons_w}
\end{equation}
In addition, the requirement to have vanishing energy density at infinity
is:
\begin{equation}
\left[r^{-m}w\right]_{r\rightarrow\infty}=0,\label{eq:bc_w}
\end{equation}
and from symmetry considerations, the energy current at the origin
must vanish as well:
\begin{equation}
\left[r^{k+d-1}w^{n}\frac{\partial w}{\partial r}\right]_{r\rightarrow0}=0.\label{eq:bc_flux}
\end{equation}

In this work, the solutions of Eq. \eqref{eq:diff_eq_1d} under the
initial and boundary conditions \eqref{eq:cons_w}-\eqref{eq:bc_flux}
will be studied in detail. We note that for $m=0$, Eq. \eqref{eq:diff_eq_1d}
represents the general non-linear diffusion equation with an inhomogeneous
diffusion coefficient of the form $D=Ar^{k}w^{n}$, which is applicable
to any other application of the diffusion equation (i.e. Ref. \cite{o1985analytical},
which deals with linear diffusion in inhomogeneous media). In contrast,
for radiation diffusion, due to the nonlinearity of the specific energy
in terms of temperature, one has in general $m\neq0$.

\section{Self-Similar solution\label{sec:Self-Similar-solution}}

Eq. \eqref{eq:diff_eq_1d} can be solved using the method dimensional
analysis \cite{buckingham1914physically,zeldovich1967physics,barenblatt1996scaling}.
This is performed in detail in Appendix \ref{sec:Dimensional-analysis}.
The result is a self similar solution whose independent dimensionless
coordinate is:
\begin{equation}
\xi=\frac{r}{\left(Q^{n}At\right)^{\frac{1}{p}}},\label{xsi_def}
\end{equation}
and the solution is given in terms of a self-similar profile:

\begin{equation}
w\left(r,t\right)=\left(\frac{Q^{2-k-m}}{\left(At\right)^{d-m}}\right)^{\frac{1}{p}}f\left(\xi\right),\label{eq:w_f}
\end{equation}
where the self similar exponent is:
\begin{equation}
p=2-k-m+\left(d-m\right)n.\label{eq:pdef}
\end{equation}
By substituting the self-similar solution \eqref{xsi_def}-\eqref{eq:w_f}
in the diffusion Eq. \eqref{eq:diff_eq_1d}, and noting that:
\begin{equation}
\frac{\partial\xi}{\partial r}=\frac{\xi}{r}=\frac{1}{\left(Q^{n}At\right)^{\frac{1}{p}}},
\end{equation}
\begin{equation}
\frac{\partial\xi}{\partial t}=-\frac{1}{p}\frac{\xi}{t},
\end{equation}
\begin{equation}
\frac{\partial w}{\partial r}=\left(\frac{Q^{2-k-m}}{\left(At\right)^{d-m}}\right)^{\frac{1}{p}}f'\left(\xi\right)\frac{\xi}{r},\label{eq:dwdr}
\end{equation}
all dimensional quantities are factored out, and a dimensionless second
order ordinary differential equation (ODE) for the self-similar solution
is obtained:

\begin{align}
\frac{d}{d\xi}\left(\xi^{k+d-1}f^{n}\left(\xi\right)\frac{df}{d\xi}+\frac{\xi^{d-m}}{p}f\left(\xi\right)\right) & =0.\label{eq:ode}
\end{align}
By substituting the self-similar variables \eqref{xsi_def}-\eqref{eq:w_f}
in Eq. \eqref{eq:cons_w}, a dimensionless conservation equation in
terms of the dimensionless variables is obtained:

\begin{equation}
\mathcal{A}_{d}\int_{0}^{\infty}f\left(\xi\right)\xi^{d-m-1}d\xi=1.\label{eq:cons_selfsim}
\end{equation}
Similarly, the boundary conditions for the energy density at infinity
(Eq. \eqref{eq:bc_w}) and the flux at the origin (Eq. \eqref{eq:bc_flux}),
are written in terms of the self-similar solution, respectively, as:
\begin{equation}
\left[\xi^{-m}f\left(\xi\right)\right]_{\xi\rightarrow\infty}=0,\label{eq:fxsi_inf}
\end{equation}
\begin{equation}
\left[\xi^{k+d-1}f^{n}\left(\xi\right)\frac{df}{d\xi}\right]_{\xi\rightarrow0}=0.\label{eq:flux_xsi_0}
\end{equation}
A direct integration of Eq. \eqref{eq:ode} gives the first order
ODE:
\begin{equation}
f^{n-1}\left(\xi\right)\frac{df}{d\xi}=-\frac{\xi^{1-k-m}}{p},\label{eq:DFDXSI}
\end{equation}
where, by virtue of the boundary condition \eqref{eq:flux_xsi_0},
the constant of the integration is set to zero. Using Eq. \eqref{eq:dwdr}
with the derivative $f'\left(\xi\right)$ taken from Eq. \eqref{eq:DFDXSI},
results in the following expression for the energy flux \eqref{eq:flux_w}:
\begin{align}
F\left(r,t\right) & =\frac{Ar^{k-1}w^{n+1}\left(r,t\right)\xi^{2-k-m}}{pf^{n}\left(\xi\right)}.\label{eq:flux_selfsim}
\end{align}

The solution of Eq. \eqref{eq:DFDXSI} takes two different forms -
for $n>0$, which corresponds to nonlinear heat conduction, and for
$n=0$, which corresponds to linear conduction. These will be analyzed
in detail in the next two sections.

\subsection{Heat front propagation}

We note that from Eq. \eqref{xsi_def}, it is evident that heat propagates
according to $r_{h}\left(t\right)\propto t^{\frac{1}{p}}$. Therefore,
in order for heat to propagate outwards, we must always have:

\begin{equation}
p>0.\label{eq:2km_constraint}
\end{equation}
This gives an upper limit for the value of $\omega$:
\begin{equation}
\omega_{\text{max}}=\frac{2+dn}{1+\lambda+\left(1-\mu\right)\left(1+n\right)}.
\end{equation}

In addition, the propagation speed is decreasing with time for $\omega<\omega_{\text{acc}}$,
constant for $\omega=\omega_{\text{acc}}$ and increasing with time
for $\omega>\omega_{\text{acc}}$, where:
\begin{equation}
\omega_{\text{acc}}=\frac{1+dn}{2+dn}\omega_{\text{max}}.
\end{equation}
It is interesting to note that since $0<\omega_{\text{acc}}<\omega_{\text{max}}$,
the well known solution to the nonlinear diffusion equation with constant
density ($\omega=0$), always has a decelerating speed of propagation,
as $r_{h}\left(t\right)\propto t^{\frac{1}{2}}$ for linear heat conduction
and $r_{h}\left(t\right)\propto t^{\frac{1}{2+dn}}$ for nonlinear
heat conduction \cite{zel1959propagation,zeldovich1967physics,barenblatt1996scaling}.
Therefore, we see that for a large enough value of $\omega$, due
to a spatial increase in the diffusion coefficient (Eq. \eqref{eq:diffpow}),
the heat propagation speed can be constant or even accelerate. This
phenomena will be demonstrated below, in Fig. \ref{fig:r_hw}.

\subsection{Relation to the strong explosion shock\label{subsec:Relation-to-the}}

The shock trajectory for a point explosion \cite{sedov1946propagation,taylor1950formation,reinicke1991point,sedov1993similarity,waxman1993second,kamm2007efficient,yalinewich2017analytic,faran2021non},
when heat conduction is negligible, behaves as $r_{shock}\left(t\right)\propto t^{\frac{2}{2+d-\omega}}$.
This $\omega$ dependent shock trajectory is often compared \cite{reinicke1991point},
for $\omega=0$, to the heat wave trajectory, $r_{h}\left(t\right)\propto t^{\frac{1}{2+dn}}$.
It is concluded that at short times, the heat wave travels faster
than the shock, so that the hydrodynamic motion is negligible and
the heat wave is considered to be ``supersonic'', while at longer
times, the heat front slows down (becomes ``subsonic''), and is
overtaken by the shock. However, it is now evident that this common
picture can change, depending on the value of $\omega$. If $\frac{1}{p}<\frac{2}{2+d-\omega}$,
we have the common case of a supersonic heat wave which is overtaken
by a shock, while for $\frac{1}{p}>\frac{2}{2+d-\omega}$, we have
a pure hydrodynamic shock at short times, which is overtaken by a
heat wave at longer times. The resulting critical spatial exponent
for this transition is:
\begin{equation}
\overline{\omega}=\frac{\left(2+d\left(2n-1\right)\right)\omega_{\text{max}}}{2\left(2+dn\right)-\omega_{\text{max}}},
\end{equation}
so that for $\omega<\overline{\omega}$ we have the common heat wave
followed by a shock behavior, while for $\omega>\overline{\omega}$,
the order is revered. For $\omega=\overline{\omega}$, both waves
propagate at the same speed, $r_{shock}\left(t\right)/r_{h}\left(t\right)=const$,
and a full rad-hydro self-similar solution may be obtained (as was
already noted in Ref. \cite{reinicke1991point} for $\omega=0$).

\subsection{Relation to Marshak waves}

The temperature profile is given by: 
\begin{equation}
T\left(r,t\right)=\left(\frac{w\left(r,t\right)}{f\rho_{0}^{1-\mu}}\right)^{\frac{1}{\beta}}.
\end{equation}
As a result, from the self-similar form in Eq. \eqref{eq:w_f}, the
temperature at the origin is given by:
\begin{equation}
T\left(0,t\right)\equiv T_{0}t^{\tau},
\end{equation}
where:
\begin{equation}
\tau=-\left(\frac{d-m}{\beta p}\right),
\end{equation}
\begin{equation}
T_{0}=\left(\frac{f\left(\xi\rightarrow0\right)}{f\rho_{0}^{1-\mu}}\left(\frac{Q^{2-k-m}}{A^{d-m}}\right)^{\frac{1}{p}}\right)^{\frac{1}{\beta}}.
\end{equation}
As a result, for planar symmetry ($d=1$), the solution presented
here is essentially a special analytic solution of the corresponding
Marshak supersonic heat wave \cite{marshak1958effect,hammer2003consistent,castor2004radiation,garnier2006self,shussman2015full,mihalas2013foundations},
with an imposed boundary temperature of the form $T_{b}\left(t\right)=T_{0}t^{\tau}$,
and a density profile of the form $\rho\left(r\right)=\rho_{0}r^{-\omega}$.

\section{Nonlinear conduction\label{sec:Nonlinear-conduction}}

\begin{table*}[t]
\centering{}%
\begin{tabular}{|c|c|c|c|c|c|c|c|}
\hline 
$ $ & $\omega\rightarrow-\infty$ & $\omega<\omega_{0}$ & $\omega=\omega_{0}$ & $\omega_{0}<\omega<\omega_{c}$ & $\omega=\omega_{c}$ & $\omega_{c}<\omega<\omega_{\text{max}}$ & $\omega\rightarrow\omega_{\text{max}}^{-}$\tabularnewline
\hline 
\hline 
$\xi_{0}$ & 1 & finite & finite & finite & finite & finite & 0\tabularnewline
\hline 
$f\left(0\right)$ & $\infty$ & finite & finite & finite & $\infty$ & $\infty$ & $\infty$\tabularnewline
\hline 
$f'\left(0\right)$ & 0 & 0 & finite & $-\infty$ & $-\infty$ & $-\infty$ & $-\infty$\tabularnewline
\hline 
\end{tabular}\caption{Behavior of the self-similar heat front coordinate $\xi_{0}$ and
the self-similar solution $f\left(0\right)$ and its derivative $f'\left(0\right)$
at the origin, for various ranges of the spatial density power $\omega$,
for nonlinear heat conduction ($n>0$).\label{tab:nonlin_behaviour}}
\end{table*}

\begin{figure}
\begin{centering}
\includegraphics[scale=0.5]{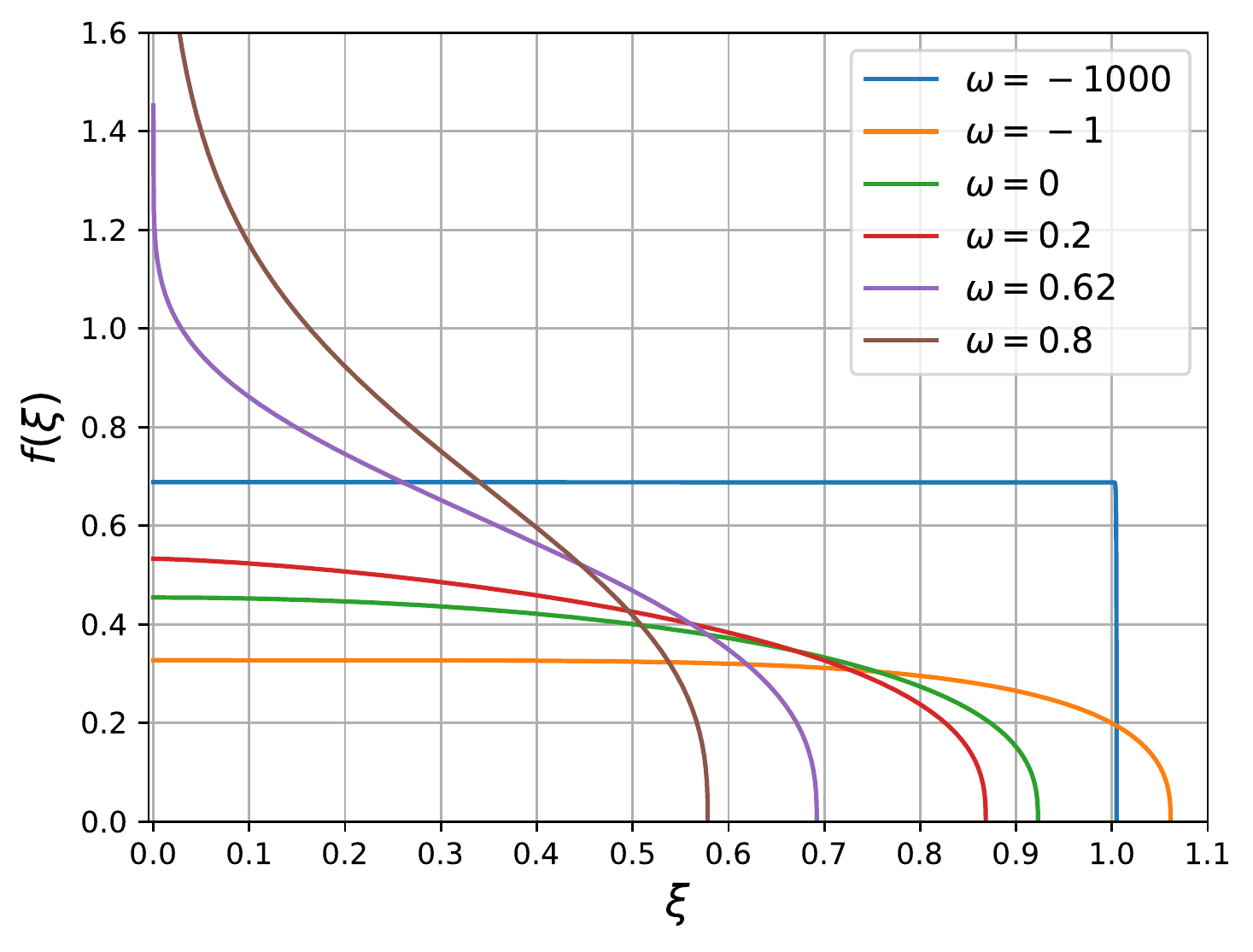}
\par\end{centering}
\caption{Various forms of the self-similar solution $f\left(\xi\right)$ for
different values of $\omega$ (listed in the legend), for the following
parameters: $d=3$ (spherical symmetry), $n=2.75$ (nonlinear conduction),
$\lambda=1$ and $\mu=0$. \label{fig:f_various}}
\end{figure}

\begin{figure}
\begin{centering}
\includegraphics[scale=0.5]{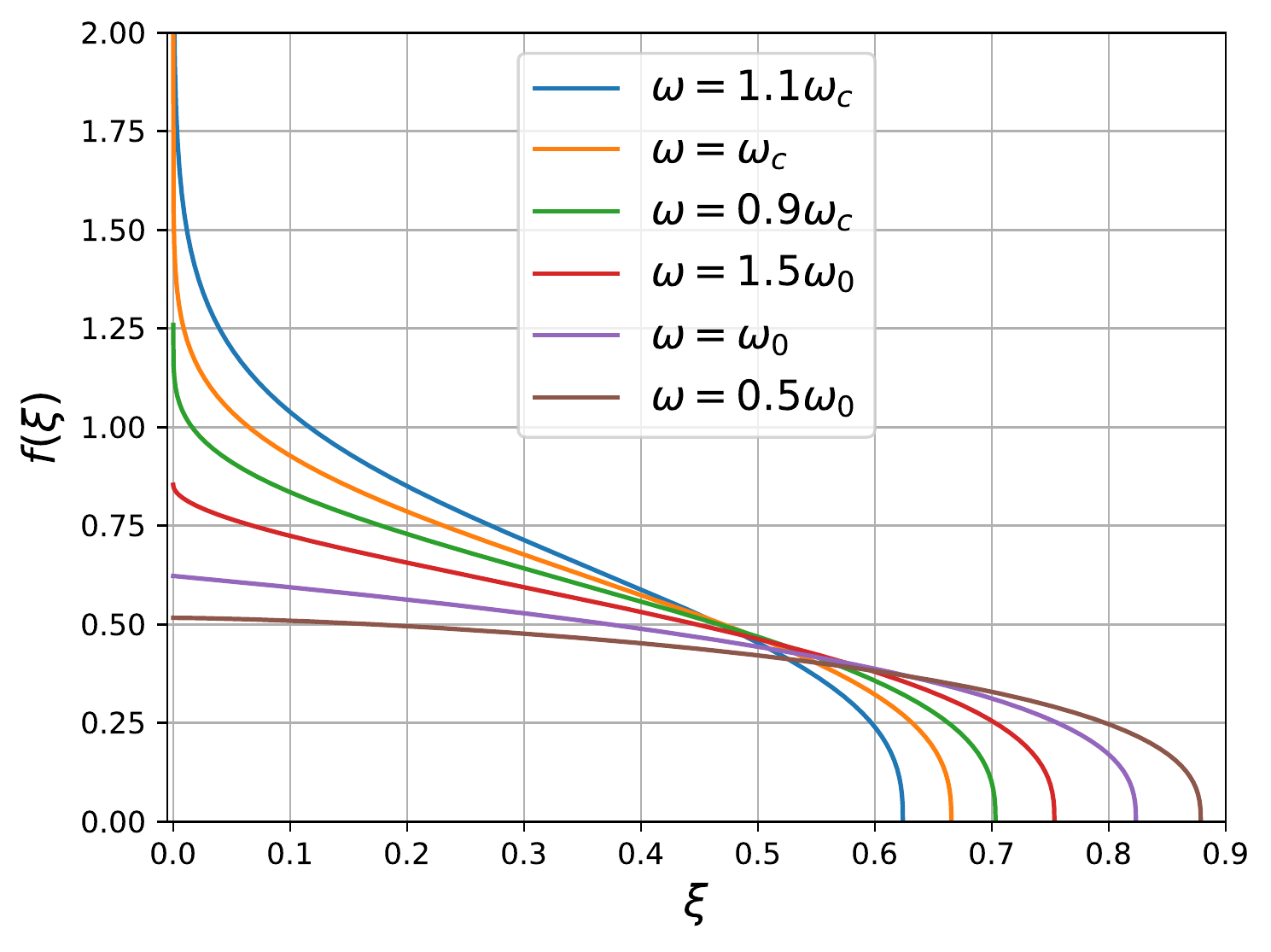}
\par\end{centering}
\begin{centering}
\includegraphics[scale=0.5]{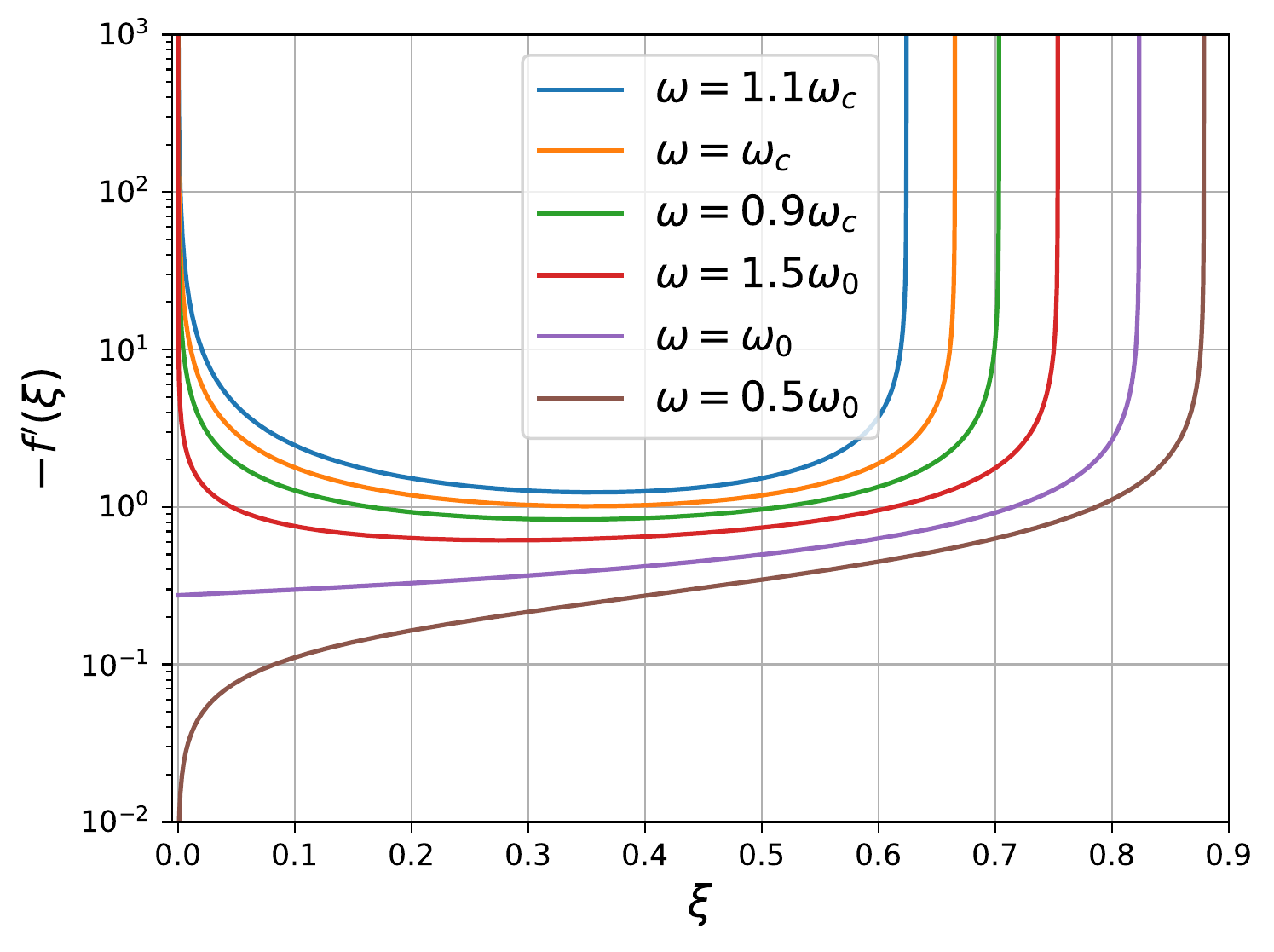}
\par\end{centering}
\caption{The self-similar solution $f\left(\xi\right)$ (upper figure) and
its derivative $-f'\left(\xi\right)$ (lower figure) for different
values of $\omega$ in the various $\omega>0$ ranges (see table \ref{tab:nonlin_behaviour}),
for the same parameters as in Fig. \ref{fig:f_various} (for which
$\omega_{\text{max}}=1.7826$, $\omega_{c}=\frac{2}{3}$ and $\omega_{0}=\frac{1}{3}$).
The solutions are plotted for the following values: $\text{\ensuremath{\omega=1.1\omega_{c}=\frac{11}{15}}}$
(blue lines), $\text{\ensuremath{\omega=\omega_{c}=\frac{2}{3}}}$
(orange lines), $\text{\ensuremath{\omega=0.9\omega_{c}=0.6}}$ (green
lines), $\text{\ensuremath{\omega=1.5\omega_{0}=0.5}}$ (red lines),
$\text{\ensuremath{\omega=\omega_{0}=\frac{1}{3}}}$ (purple lines)
and $\text{\ensuremath{\omega=0.5\omega_{c}=\frac{1}{6}}}$ (brown
lines). \label{fig:f_df}}
\end{figure}

\begin{figure}
\begin{centering}
\includegraphics[scale=0.5]{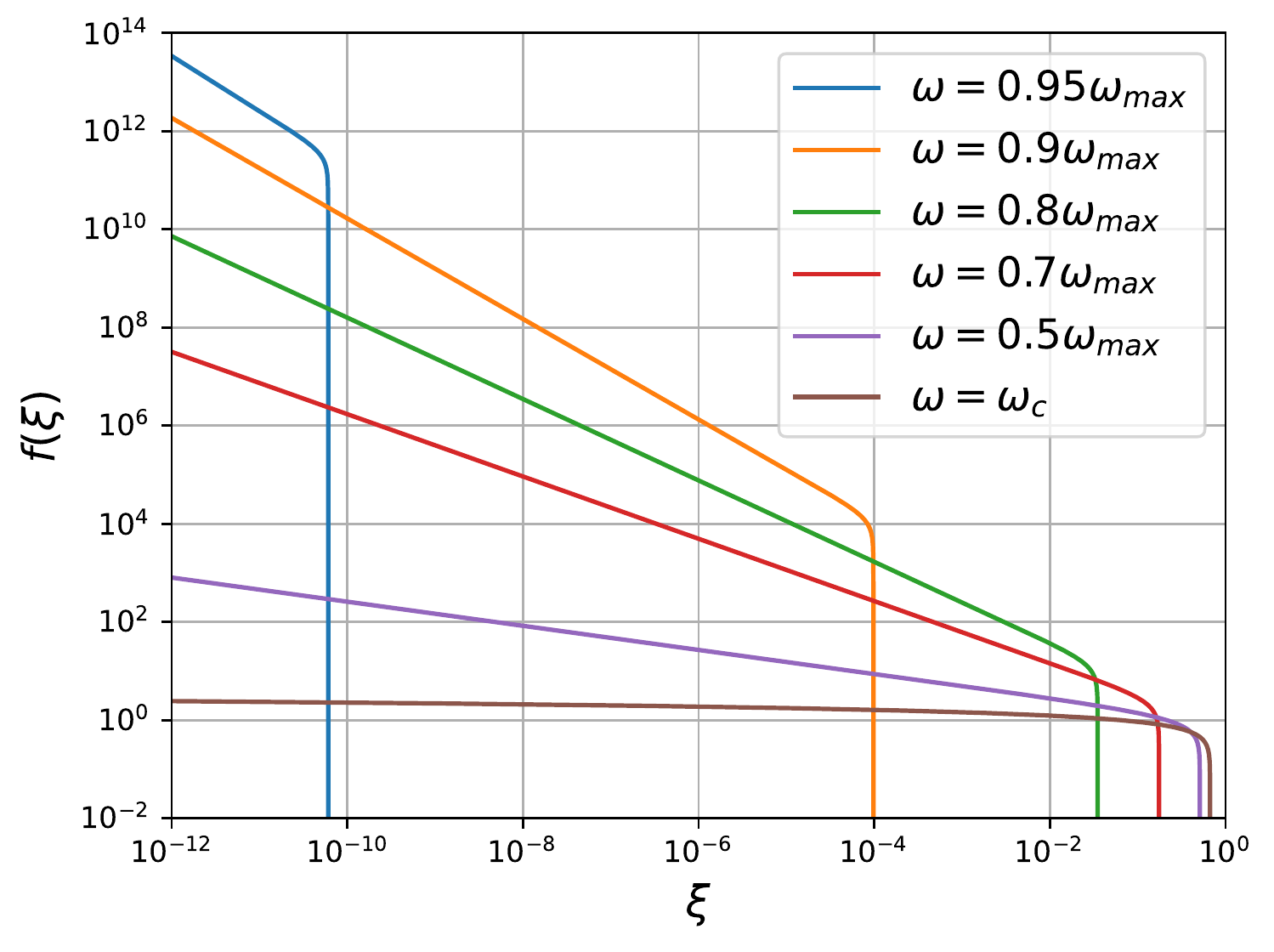}
\par\end{centering}
\caption{The self-similar solution $f\left(\xi\right)$ for different values
of $\omega$ near $\omega_{\text{max}}$, for the same parameters
of Fig. \ref{fig:f_various} ($\omega_{\text{max}}=1.7826$). The
solutions are plotted for the following values: $\text{\ensuremath{\omega=0.95\omega_{\text{max}}=1.693}}$
(blue line), $\text{\ensuremath{\omega=0.9\omega_{\text{max}}=1.604}}$
(orange line), $\text{\ensuremath{\omega=0.8\omega_{\text{max}}=1.426}}$
(green line), $\text{\ensuremath{\omega=0.7\omega_{\text{max}}=1.248}}$
(red line), $\text{\ensuremath{\omega=0.5\omega_{\text{max}}=0.891}}$
(purple line) and $\text{\ensuremath{\omega=\omega_{c}=\frac{2}{3}}}$
(brown line). \label{fig:f_omega_max}}
\end{figure}

\begin{figure}
\begin{centering}
\includegraphics[scale=0.5]{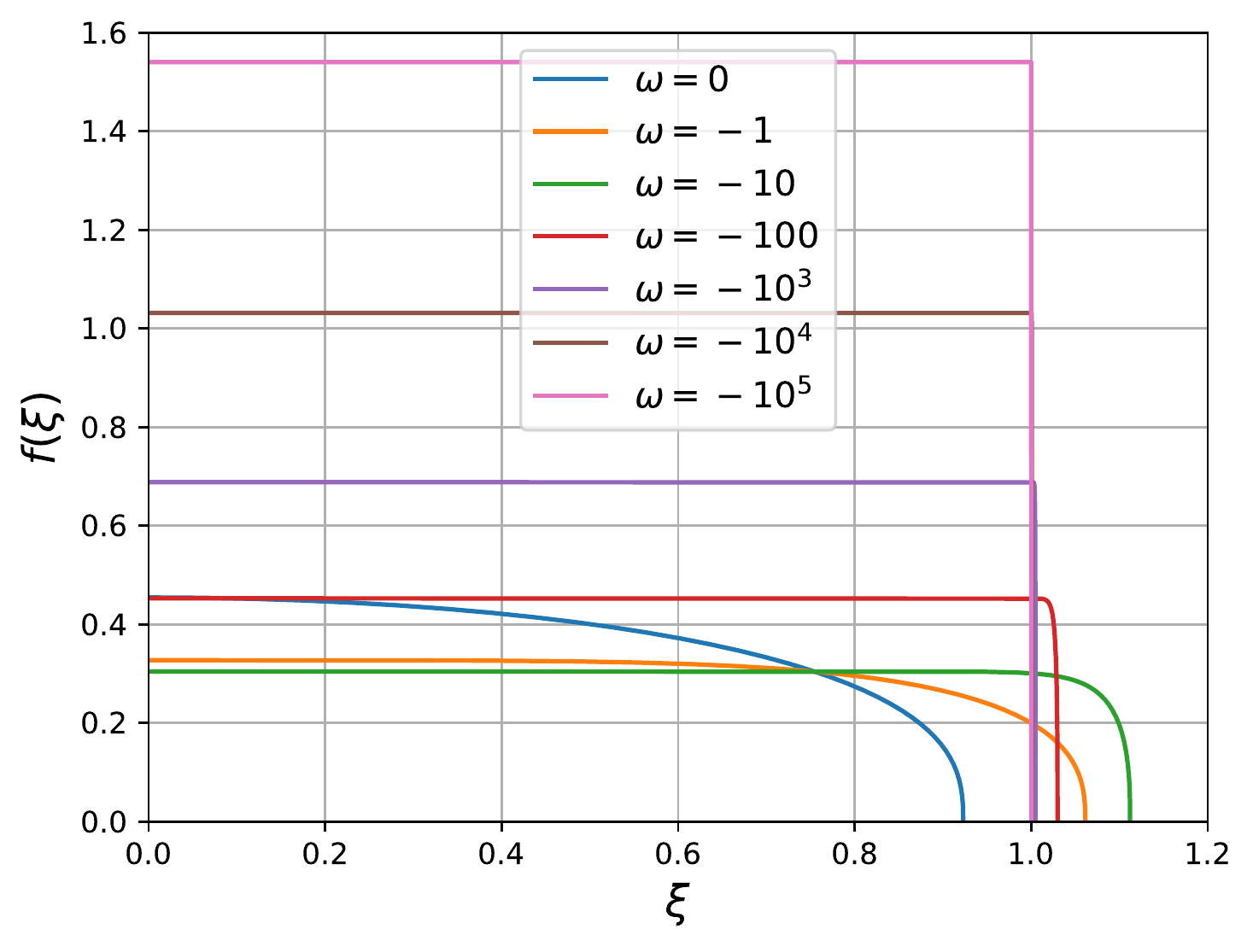}
\par\end{centering}
\caption{The self-similar solution $f\left(\xi\right)$ for decreasing values
of $\omega$ (listed in the legend), for the same parameters of Fig.
\ref{fig:f_various} (nonlinear conduction). \label{fig:f_omega_inf}}
\end{figure}

\begin{figure*}[t]
\begin{centering}
\includegraphics[scale=0.5]{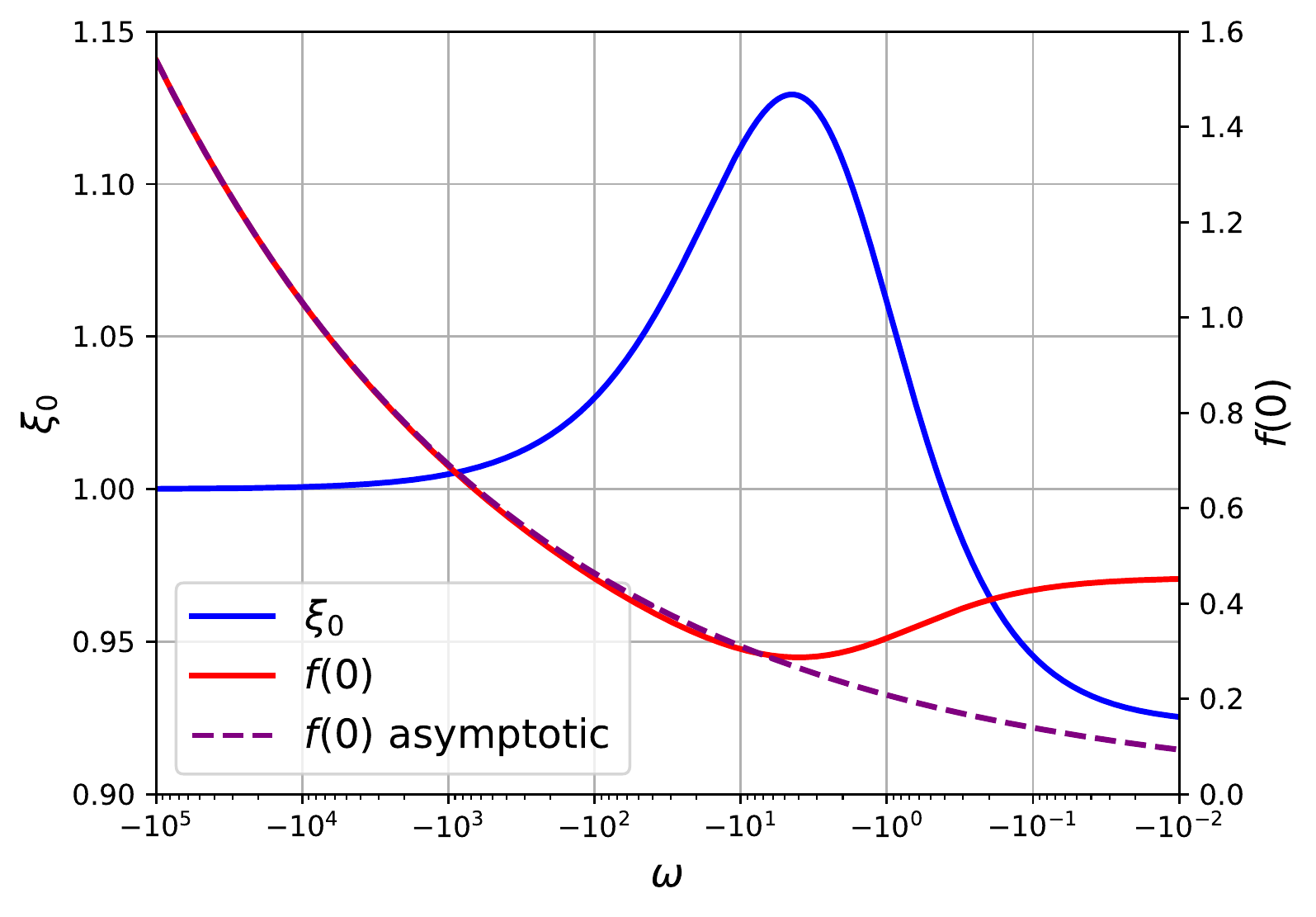}\includegraphics[scale=0.5]{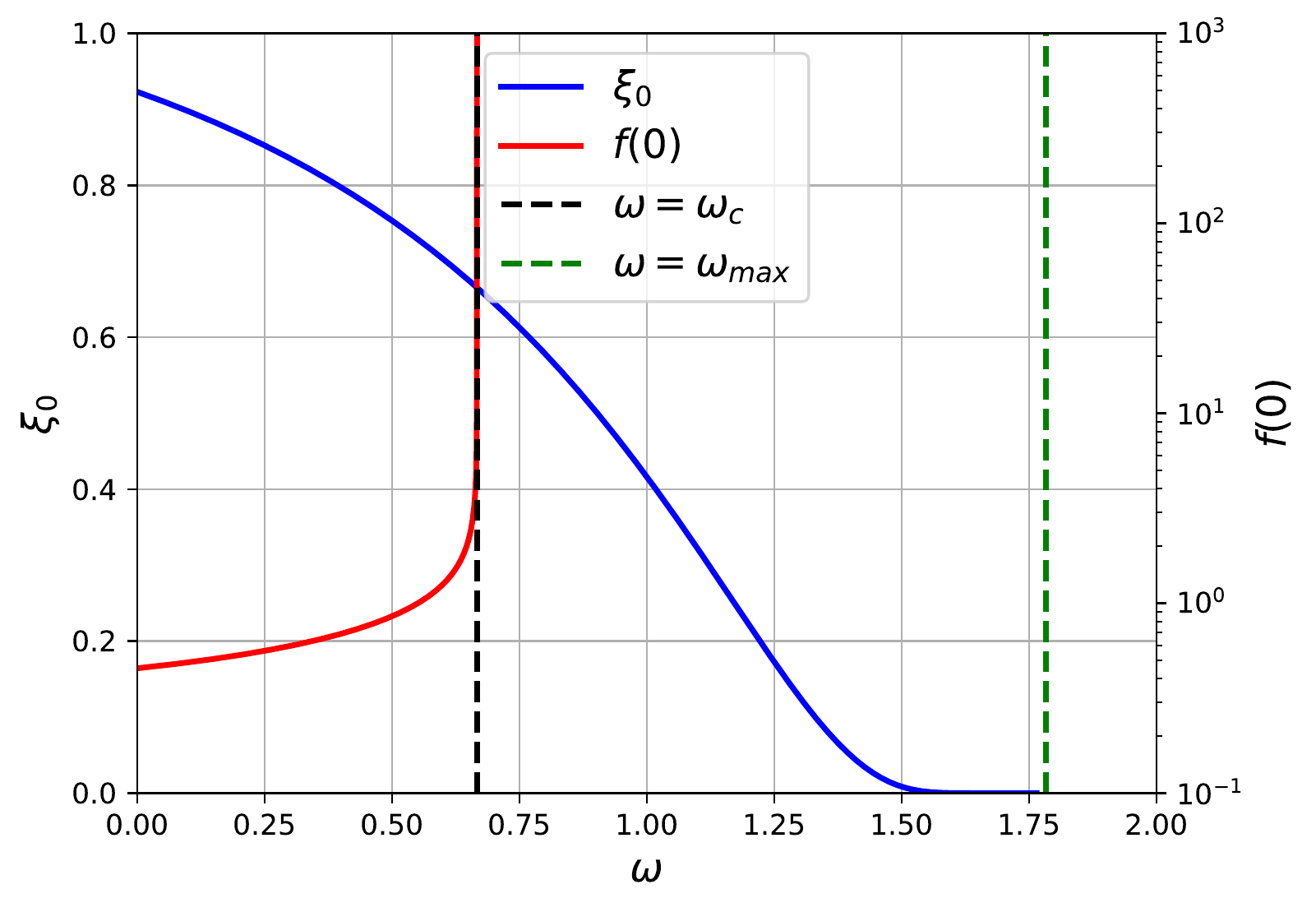}
\par\end{centering}
\caption{The values of the self-similar heat front coordinate $\xi_{0}$ (in
blue, on the left y axes) and $f\left(0\right)$ (in red, on the right
y axes), as a function of $\omega$. The problem's parameters are
the same as in Fig. \ref{fig:f_various} (nonlinear conduction). Negative
$\omega$ values, in a wide range, are plotted on the left figure,
while positive values are plotted on the right figure. The asymptotic
approximation (Eq. \eqref{eq:fste_nonlin}) of $f\left(0\right)$
for $\omega\rightarrow-\infty$ is given in the purple dashed line.
The special limiting values $\omega_{c}$ and $\omega_{\text{max}}$
(see table \ref{tab:nonlin_behaviour}), are given by the black and
green vertical lines, respectively. \label{fig:f0_xsi0}}
\end{figure*}

Assuming $n>0$, and employing the boundary condition at infinity
(see Eq. \eqref{eq:fxsi_inf}), the ODE in Eq. \eqref{eq:DFDXSI}
has a simple analytic solution of the form:

\begin{equation}
f\left(\xi\right)=\begin{cases}
\left(\frac{n\left(\xi_{0}^{2-k-m}-\xi^{2-k-m}\right)}{p\left(2-k-m\right)}\right)^{\frac{1}{n}}, & \xi<\xi_{0}\\
0, & \xi>\xi_{0}
\end{cases}\label{eq:fxsi_sol}
\end{equation}
where $\xi_{0}$ is a constant, which represents the self similar
coordinate of the heat wave position. Hence, the heat wave position
as a function of time is given by:

\begin{equation}
r_{h}\left(t\right)=\xi_{0}\left(Q^{n}At\right)^{\frac{1}{p}}.\label{eq:hw_nonlin}
\end{equation}
The value of $\xi_{0}$ can be found by substituting the solution
\eqref{eq:fxsi_sol} in the energy conservation constraint \eqref{eq:cons_selfsim}.
The resulting value of $\xi_{0}$, which is derived in detail in Appendix
\ref{app:xsi0}, is:

\begin{equation}
\xi_{0}=\left(\frac{p\left|2-k-m\right|^{n+1}}{n\mathcal{A}_{d}^{n}\mathcal{B}^{n}\left(l,\frac{1}{n}+1\right)}\right)^{\frac{1}{p}}\label{eq:xsi0_final}
\end{equation}
where the first beta function argument is:
\begin{equation}
l=\begin{cases}
\frac{d-m}{2-k-m}, & 2-k-m>0\\
-\frac{1}{n}-\frac{d-m}{2-k-m}, & 2-k-m<0
\end{cases}\label{eq:l_xsi_0}
\end{equation}
The solution in Eq. \eqref{eq:fxsi_sol} is valid only under the constraint
$d-m>0$ and \eqref{eq:2km_constraint} (which is equivalent to $\omega<\omega_{\text{max}}$).

It is evident that for the constant density case ($\omega=k=m=0$),
the solution \eqref{eq:fxsi_sol} is reduced to the well known nonlinear
heat wave \cite{zeldovich1967physics,zel1959propagation,barenblatt1996scaling,mihalas2013foundations},
which has a sharp front at $\xi\rightarrow\xi_{0}$, and takes the
form $f\left(\xi\right)\propto\left(\xi_{0}^{2}-\xi^{2}\right)^{\frac{1}{n}}$,
for which $f'\left(0\right)=0$, and the solution approaches a finite
constant value near the origin. However, for a non-homogeneous media
($\omega\neq0$, which results in $k,m\neq0$), which is considered
in this work, this form is generalized to $f\left(\xi\right)\propto\left(\xi_{0}^{2-k-m}-\xi^{2-k-m}\right)^{\frac{1}{n}}$,
which may differ qualitatively from the traditional $\omega=0$ solution.
In fact, as demonstrated in Fig. \ref{fig:f_various}, several different
qualitative forms take place, according to the value of $\omega$.
It is seen that: (i) for large enough values of $\omega$, the solution
diverges at the origin, (ii) for smaller values of $\omega$ the solution
is finite at the origin but its derivative diverges, (iii) for smaller
values of $\omega$ the solution approaches a constant near the origin
(a familiar property of the solution for $\omega=0$), and finally
(iv) for large negative values of $\omega$ the solution becomes steeper
and approaches a step function, where the step value diverges as $\omega\rightarrow-\infty$.
These different forms shown in Fig. \ref{fig:f_various}, will be
analyzed in detail below, and are summarized in table \ref{tab:nonlin_behaviour}.

\subsection{Analysis of various solution forms for different values of $\omega$}

First we note that since $f'\left(\xi\rightarrow0\right)\propto\xi^{1-k-m}$.
For $1-k-m<0$ one has $f'\left(0\right)\rightarrow-\infty$, so that
the solution slope diverges near the origin. This condition is equivalent
to $\omega>\omega_{0}$, where:
\begin{equation}
\omega_{0}=\frac{1}{2+\lambda-\mu}.
\end{equation}
For the special value $\omega=\omega_{0}$, the value of $f'\left(0\right)$
is finite and can be obtained from Eq. \eqref{eq:DFDXSI}, so that
we have:

\begin{equation}
f'\left(0\right)=\begin{cases}
-\infty, & \omega>\omega_{0}\\
-\frac{1}{pf^{n-1}\left(0\right)}, & \omega=\omega_{0}\\
0, & \omega<\omega_{0}
\end{cases}\label{eq:ftag_0}
\end{equation}
Moreover, the solution \eqref{eq:fxsi_sol} is valid even for $2-k-m<0$
(provided that Eq. \eqref{eq:2km_constraint} holds), for which the
self similar profile itself diverges at the origin. The latter condition
is equivalent to $\omega>\omega_{c}$, where:
\begin{equation}
\omega_{c}=2\omega_{0}.
\end{equation}
We also note that for the special value $\omega=\omega_{c}$, the
solution in equations \eqref{eq:fxsi_sol},\eqref{eq:xsi0_final}
and \eqref{eq:l_xsi_0} cannot be used directly (since $2-k-m=0$).
However, the ODE \eqref{eq:DFDXSI} can be solved independently for
this case, as done in detail in Appendix \ref{sec:Analytic-solution-foroemgac}.
The resulting solution takes the marginal form:

\begin{equation}
f\left(\xi\right)=\begin{cases}
\left[\frac{1}{d-m}\ln\left(\frac{\xi_{0}}{\xi}\right)\right]^{\frac{1}{n}}, & \xi<\xi_{0}\\
0, & \xi>\xi_{0}
\end{cases}\label{eq:omega_c_sol}
\end{equation}
where:

\begin{equation}
\xi_{0}=\left[\frac{\left(d-m\right)^{1+\frac{2}{n}}}{\mathcal{A}_{d}\Gamma\left(1+\frac{1}{n}\right)}\right]^{\frac{1}{d-m}}.\label{eq:xsi0_omegac}
\end{equation}

The solution $f\left(\xi\right)$ and its derivative $f'\left(\xi\right)$
are shown in Fig. \ref{fig:f_df} for various values of $\omega$
(see table \ref{tab:nonlin_behaviour}): $\omega_{c}<\omega<\omega_{\text{max}}$
(diverging solution and derivative at the origin), $\omega=\omega_{c}$
(marginal form \eqref{eq:omega_c_sol}), $\omega_{0}<\omega<\omega_{c}$
(finite solution and diverging derivative at the origin), $\omega=\omega_{0}$
(finite solution and derivative at the origin) and $\omega<\omega_{0}$
(derivative is zero at the origin). 

Fig. \ref{fig:f_omega_max} shows various solutions for $\omega\geq\omega_{c}$
close to $\omega_{\text{max}}$. The divergence of these solutions
near the origin is evident, as well as the fact that:
\begin{equation}
\lim_{\omega\rightarrow\omega_{\text{max}}^{-}}\xi_{0}=0,
\end{equation}
which is straightforward to show directly from Eq. \eqref{eq:xsi0_final},
or from the fact that energy conservation \eqref{eq:cons_selfsim}
must hold. This is also shown in Fig. \ref{fig:f0_xsi0}, where the
values of $\xi_{0}$ and $f\left(0\right)$ are plotted as a function
of $\omega$.

Finally, we consider the interesting limiting solution for $\omega\rightarrow-\infty$.
First we note that it is straightforward to calculate that limit of
Eq. \eqref{eq:xsi0_final}, which is:
\begin{equation}
\lim_{\omega\rightarrow-\infty}\xi_{0}=1,\label{eq:xsi_0_omega_in_nonlion}
\end{equation}
and to show that the resulting limiting solution \eqref{eq:fxsi_sol}
has a step function form:
\begin{equation}
\lim_{\omega\rightarrow-\infty}f\left(\xi\right)=\begin{cases}
B\left|\omega\right|^{b}, & \xi<1\\
0, & \text{else}
\end{cases}\label{eq:fste_nonlin}
\end{equation}
where:
\begin{equation}
b=\frac{\lambda+\mu}{2+dn}\omega_{\text{max}},\label{eq:bdef}
\end{equation}
and:
\begin{equation}
B=\frac{\left(\omega_{0}^{1-q\left(n+1\right)}\left(\frac{n\omega_{\text{max}}}{2+dn}\right)^{1-q}\right)^{\frac{1}{n}}}{\left(\mathcal{A}_{d}\mathcal{B}\left(\bar{l},\frac{1}{n}+1\right)\right)^{q}},
\end{equation}
where:

\begin{equation}
q=\frac{\omega_{\text{max}}}{\omega_{0}\left(2+dn\right)},
\end{equation}
and the $\omega\rightarrow-\infty$ limit of Eq. \eqref{eq:l_xsi_0}
is:
\begin{equation}
\bar{l}=\left(1-\mu\right)\omega_{0}.
\end{equation}
We note that since $b>0$, the step function diverges as $\omega\rightarrow-\infty$,
which is of course necessary in order for the energy conservation
constraint \eqref{eq:cons_selfsim} to hold. The steepening of the
solution for increasing values of $\omega$ as well as the divergence
of the step value is demonstrated in Fig. \ref{fig:f_omega_inf}.
The limit \eqref{eq:xsi_0_omega_in_nonlion} and the asymptotic form
\eqref{eq:fste_nonlin} are demonstrated in Fig. \ref{fig:f0_xsi0}.

Figs \ref{fig:f_various}-\ref{fig:f0_xsi0} are calculated for the
following case study: $d=3$ (spherical symmetry), $n=2.75$, $\lambda=1$
and $\mu=0$, for which $\omega_{\text{max}}=1.7826$, $\omega_{c}=\frac{2}{3}$
and $\omega_{0}=\frac{1}{3}$.

\section{linear conduction\label{sec:linear-conduction}}

Assuming $n=0$, which corresponds to the linear conduction case,
the solution of Eq. \eqref{eq:DFDXSI}, has a super-Gaussian form:
\begin{equation}
f\left(\xi\right)=f_{0}\exp\left(-\frac{\xi^{2-k-m}}{\left(2-k-m\right)^{2}}\right).\label{eq:fxsi_anal_linear}
\end{equation}
As in the nonlinear case, the constant $f_{0}$ is obtained from the
energy conservation constraint \eqref{eq:cons_selfsim}:

\begin{equation}
f_{0}=\frac{1}{\mathcal{A}_{d}\Gamma\left(\frac{d-m}{2-k-m}\right)\left(2-k-m\right)^{\frac{2\left(d-m\right)+k+m-2}{2-k-m}}},\label{eq:f0_linear}
\end{equation}
where we have employed the well known identity:
\begin{equation}
\int_{0}^{\infty}x^{b}\exp\left(-cx^{a}\right)dx=\frac{\Gamma\left(\frac{b+1}{a}\right)}{ac^{\frac{b+1}{a}}}.
\end{equation}
As in the nonlinear conduction case, the solution in equations \eqref{eq:fxsi_anal_linear}-\eqref{eq:f0_linear}
is valid only under the constraint \eqref{eq:2km_constraint}, which
is now equivalent to $\omega<\omega_{\text{max}}=\omega_{c}$. As
a result, this solution also obeys the boundary condition \eqref{eq:fxsi_inf}.

It is evident that for the constant density case ($\omega=k=m=0$),
the solution \eqref{eq:fxsi_anal_linear}-\eqref{eq:f0_linear} is
reduced to the well known Gaussian solution to the linear diffusion
equation:

\begin{equation}
f\left(\xi\right)=\frac{1}{2^{d}\pi^{\frac{d}{2}}}e^{-\frac{\xi^{2}}{4}},\ \xi=\frac{r}{\left(At\right)^{\frac{1}{2}}},
\end{equation}
for which $f'\left(0\right)=0$, and the solution approaches a finite
constant value near the origin. However, similarly to the non-linear
conduction case that was considered in the previous section, for a
non-homogeneous media ($\omega\neq0$, which results in $k,m\neq0$),
this form is generalized to $f\left(\xi\right)\propto e^{-\xi^{2-k-m}}$,
which may differ qualitatively from the traditional $\omega=0$ solution.
This is demonstrated in Fig. \ref{fig:f_various_lin}, where several
different qualitative forms are shown for different values of $\omega$.
These different forms will be analyzed in detail below, and are summarized
in table \ref{tab:lin_behaviour}.

\subsection{Analysis of various solution forms for different values of $\omega$}

As in the nonlinear conduction case, Eq. \eqref{eq:fxsi_anal_linear}
gives $f'\left(\xi\rightarrow0\right)\propto\xi^{1-k-m}$, so that
behavior of the derivative near the origin is the same as in the nonlinear
case, and specifically, Eq. \eqref{eq:ftag_0} is applicable for $n=0$
as well.

Similarly, Eq. \eqref{eq:fxsi_anal_linear} gives $f\left(\xi\rightarrow0\right)\propto\xi^{2-k-m}$,
but since $\omega<\omega_{\text{max}}=\omega_{c}$ for $n=0$ (equivalent
to $2-k-m>0$), we see that the solution is always finite at the origin
for linear conduction. The solution $f\left(\xi\right)$ and its derivative
$f'\left(\xi\right)$ are shown in Fig. \ref{fig:f_df_lin} for various
values of $\omega$ (see table \ref{tab:lin_behaviour}): $\omega_{0}<\omega<\omega_{c}$
(finite solution and diverging derivative at the origin), $\omega=\omega_{0}$
(finite solution and derivative at the origin) and $\omega<\omega_{0}$
(derivative is zero at the origin). 

Fig. \ref{fig:f_omega_max_lin} shows various solutions for $\omega\geq\omega_{0}$
close to $\omega_{\text{max}}$. It is seen that these solutions are
more concentrated near the origin for $\omega$ closer to $\omega_{\text{max}}$
and that:
\begin{equation}
\lim_{\omega\rightarrow\omega_{\text{max}}^{-}}f\left(0\right)=\infty,
\end{equation}
as can be shown directly from Eq. \eqref{eq:f0_linear}, or from the
fact that energy conservation \eqref{eq:cons_selfsim} must hold.
This is also shown in Fig. \ref{fig:f0_xsi0_lin}, where the values
of $\xi_{0}$ and $f\left(0\right)$ are plotted as a function of
$\omega$.

Finally, we consider the limiting solution for $\omega\rightarrow-\infty$.
It is straightforward to show that the resulting limiting solution
\eqref{eq:fxsi_anal_linear}, has the same step function form as in
the nonlinear case, Eq. \eqref{eq:fste_nonlin}, with $b=\omega_{0}\left(\lambda+\mu\right)$
(which is the same as Eq. \eqref{eq:bdef} for $n=0$), and:
\begin{equation}
B=\frac{1}{\mathcal{A}_{d}\omega_{0}^{b}\Gamma\left(\left(1-\mu\right)\omega_{0}\right)}.\label{eq:Blin}
\end{equation}
As noted for the nonlinear case, the step function diverges as $\omega\rightarrow-\infty$,
which is of course necessary in order for the energy conservation
constraint \eqref{eq:cons_selfsim} to hold. The steepening of the
solution for increasing values of $\omega$ as well as the divergence
of the step value is demonstrated in Fig. \ref{fig:f_omega_in_lin}.
The asymptotic form \eqref{eq:fste_nonlin} (using Eq. \eqref{eq:Blin})
is shown in Fig. \ref{fig:f0_xsi0_lin} and agrees well with the solution
for $\omega\rightarrow-\infty$.

\begin{figure}
\begin{centering}
\includegraphics[scale=0.5]{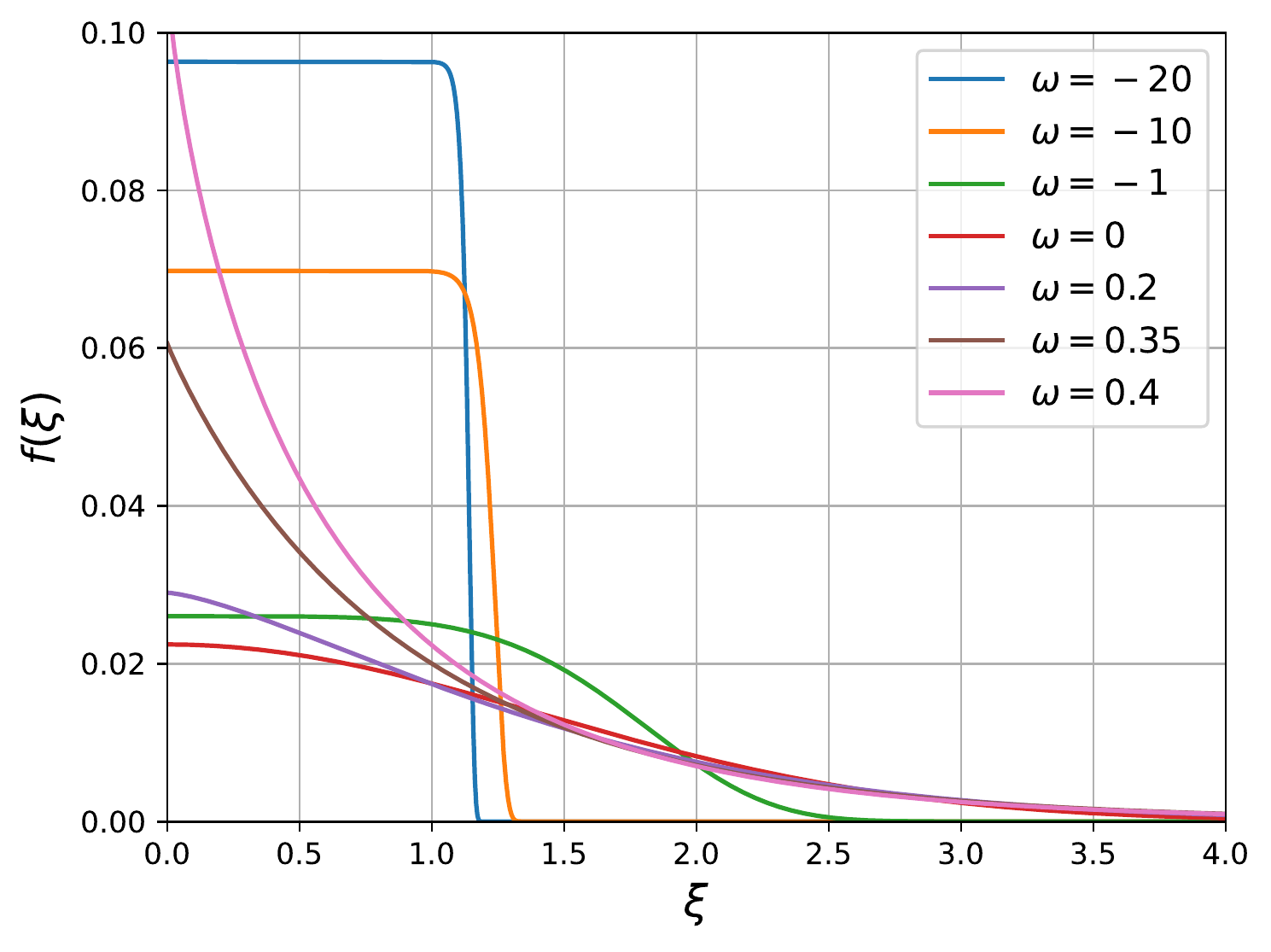}
\par\end{centering}
\caption{Various forms of the self-similar solution $f\left(\xi\right)$ for
different values of $\omega$ (listed in the legend), for the following
parameters: $d=3$ (spherical symmetry), $n=0$ (linear conduction),
$\lambda=1$ and $\mu=0$. \label{fig:f_various_lin}}
\end{figure}

\begin{figure}
\begin{centering}
\includegraphics[scale=0.5]{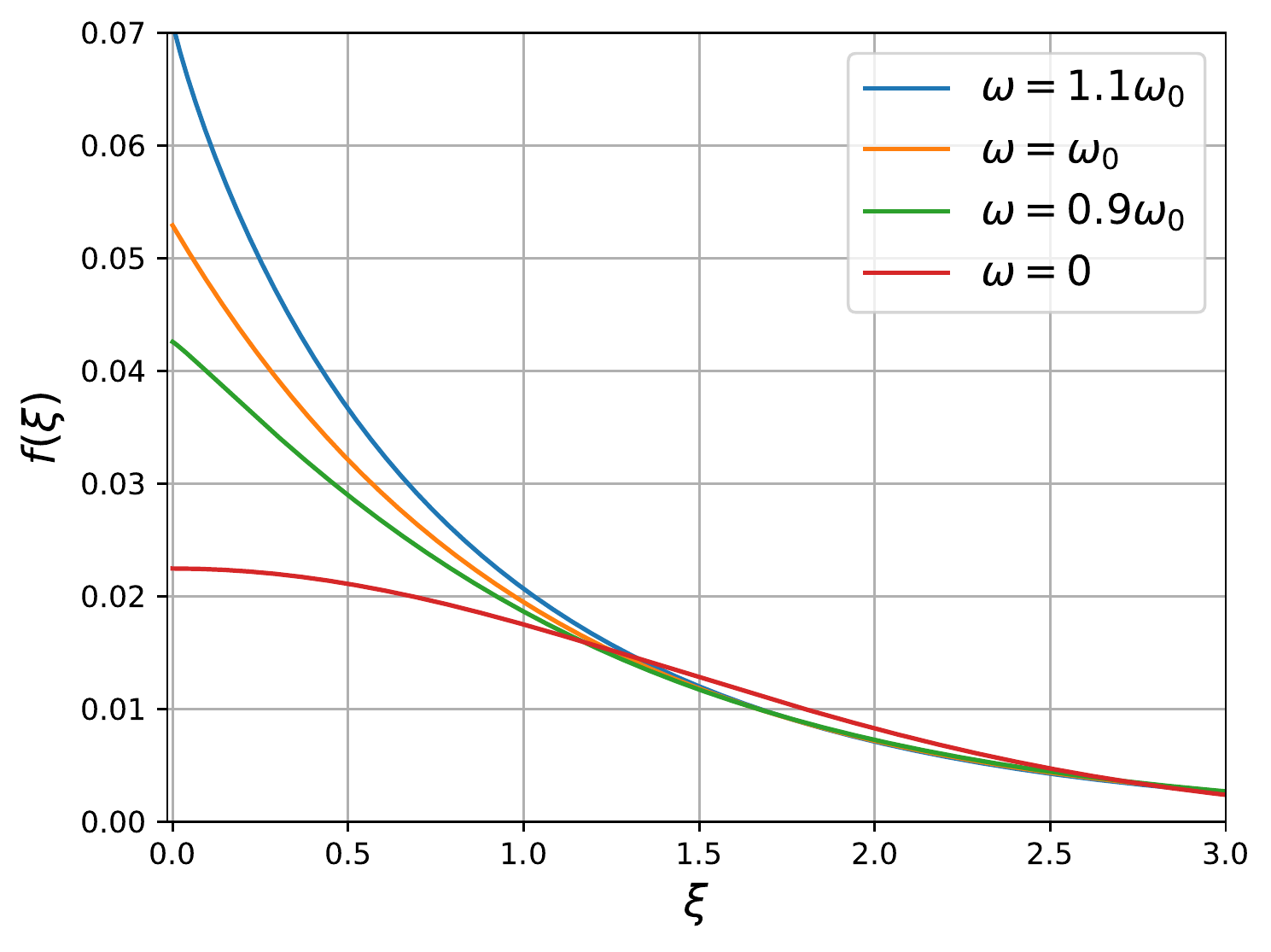}
\par\end{centering}
\begin{centering}
\includegraphics[scale=0.5]{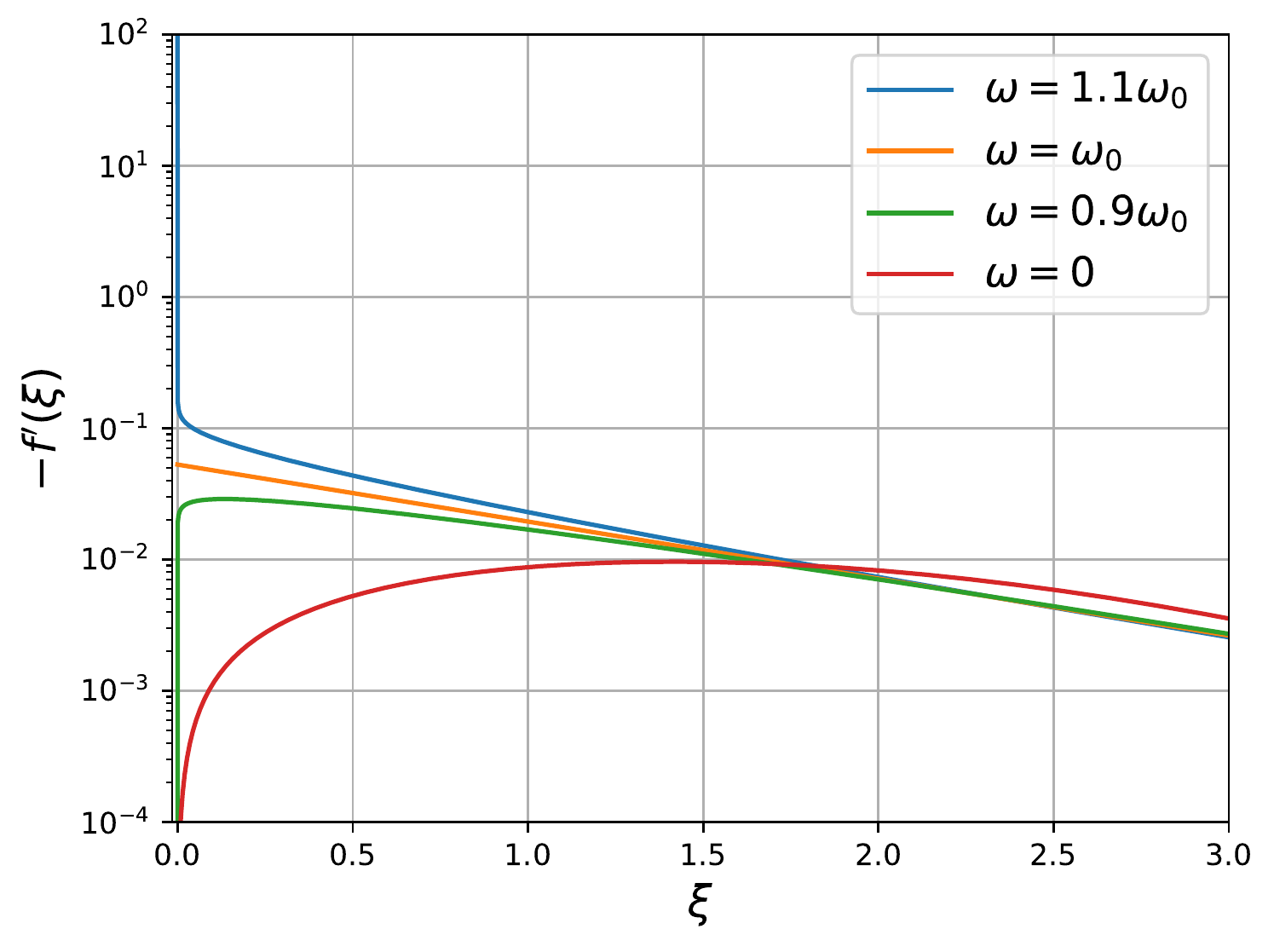}
\par\end{centering}
\caption{The self-similar solution $f\left(\xi\right)$ (upper figure) and
its derivative $-f'\left(\xi\right)$ (lower figure) for different
values of $\omega$ (see table \ref{tab:lin_behaviour}), for the
same parameters of Fig. \ref{fig:f_various_lin} (for which $\omega_{c}=\omega_{\text{max}}=\frac{2}{3}$
and $\omega_{0}=\frac{1}{3}$). The solutions are plotted for the
following values: $\text{\ensuremath{\omega=1.1\omega_{0}=\frac{11}{30}}}$
(blue lines), $\text{\ensuremath{\omega=\omega_{0}=\frac{1}{3}}}$
(orange lines), $\text{\ensuremath{\omega=0.9\omega_{0}=0.3}}$ (green
lines), $\text{\ensuremath{\omega=0}}$ (red lines). \label{fig:f_df_lin}}
\end{figure}

\begin{figure}
\begin{centering}
\includegraphics[scale=0.5]{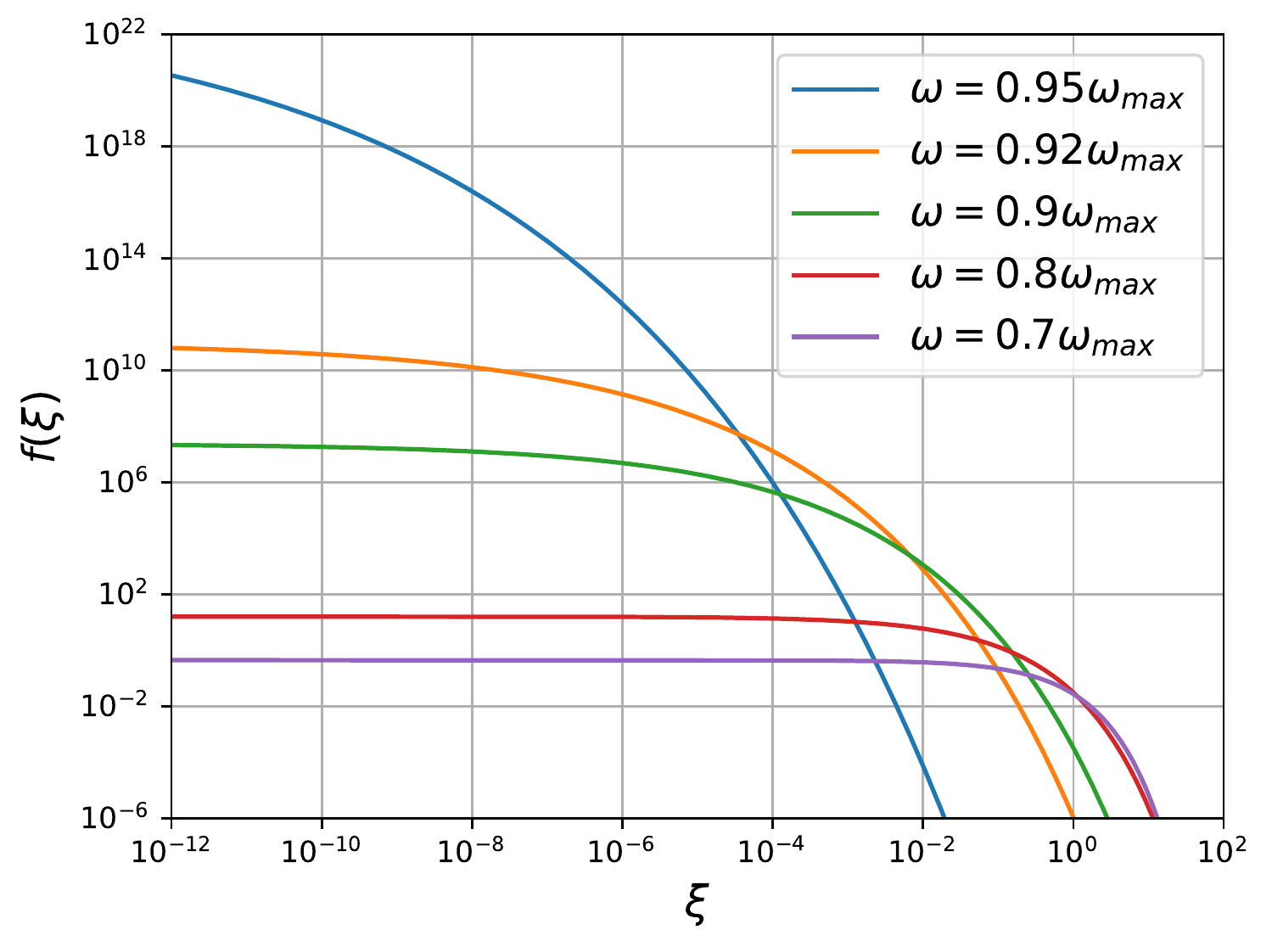}
\par\end{centering}
\caption{The self-similar solution $f\left(\xi\right)$ for different values
of $\omega$ near $\omega_{\text{max}}$, for the same parameters
of Fig. \ref{fig:f_various_lin} ($\omega_{\text{max}}=\frac{2}{3}$).
The solutions are plotted for the following values: $\text{\ensuremath{\omega=0.95\omega_{\text{max}}=0.633}}$
(blue line), $\text{\ensuremath{\omega=0.92\omega_{\text{max}}=0.613}}$
(orange line), $\text{\ensuremath{\omega=0.9\omega_{\text{max}}=0.6}}$
(green line), $\text{\ensuremath{\omega=0.8\omega_{\text{max}}=0.533}}$
(red line) and $\text{\ensuremath{\omega=0.7\omega_{\text{max}}=0.466}}$
(purple line). \label{fig:f_omega_max_lin}}
\end{figure}

\begin{figure}
\begin{centering}
\includegraphics[scale=0.5]{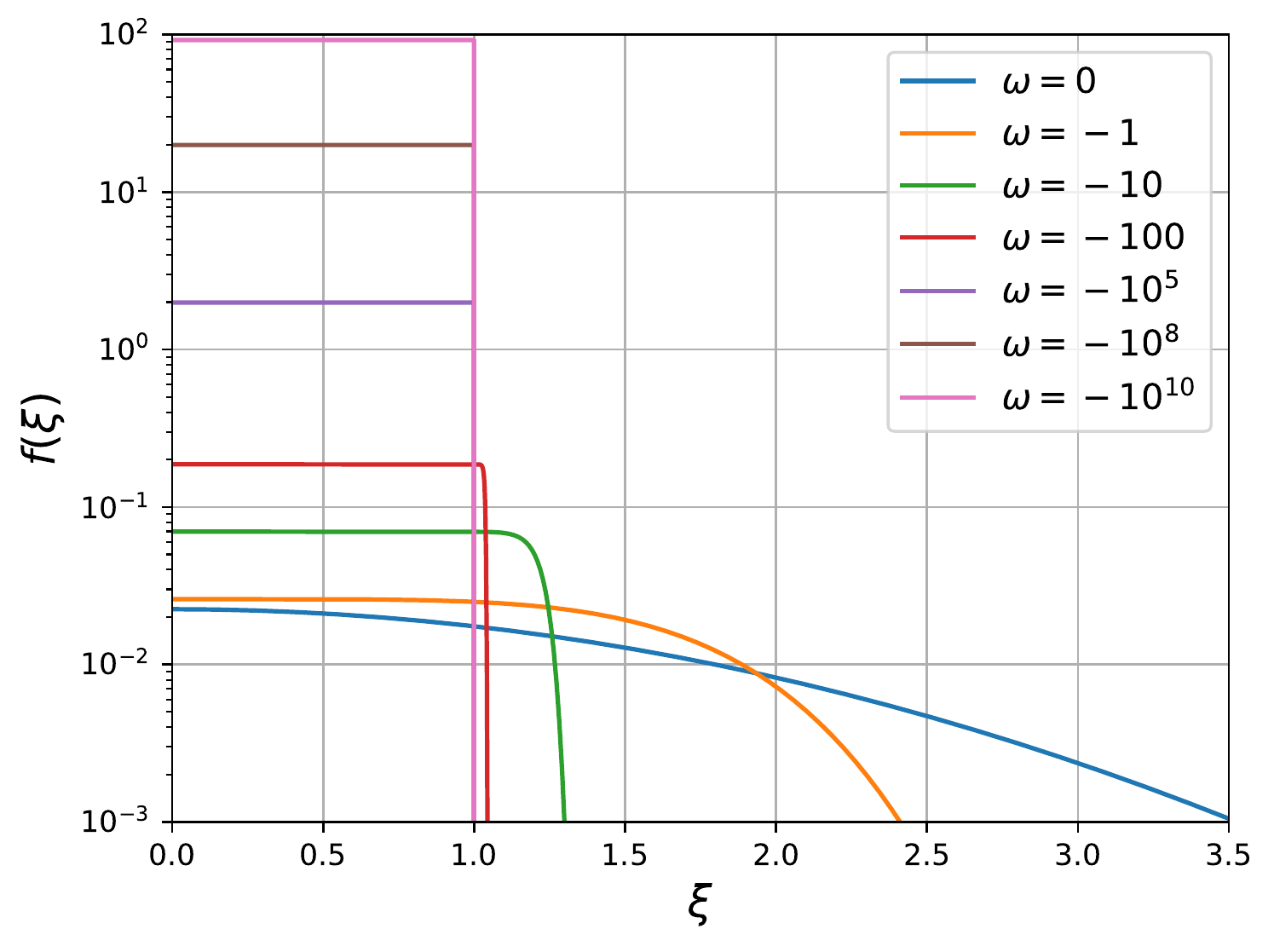}
\par\end{centering}
\caption{The self-similar solution $f\left(\xi\right)$ for increasing values
of $\omega$, for the same parameters of Fig. \ref{fig:f_various_lin}
(linear conduction). \label{fig:f_omega_in_lin}}
\end{figure}

\begin{figure*}[t]
\begin{centering}
\includegraphics[scale=0.5]{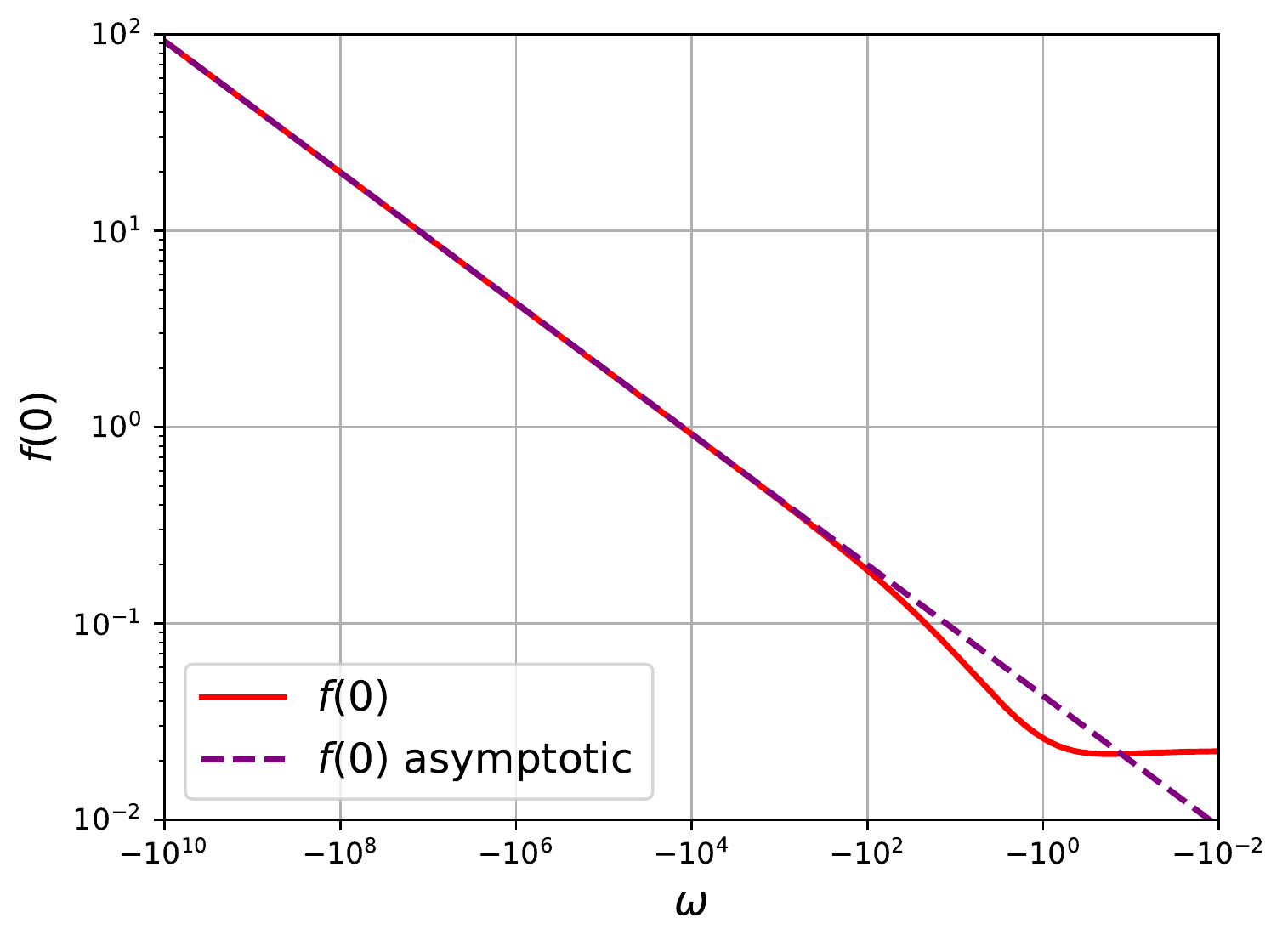}\includegraphics[scale=0.5]{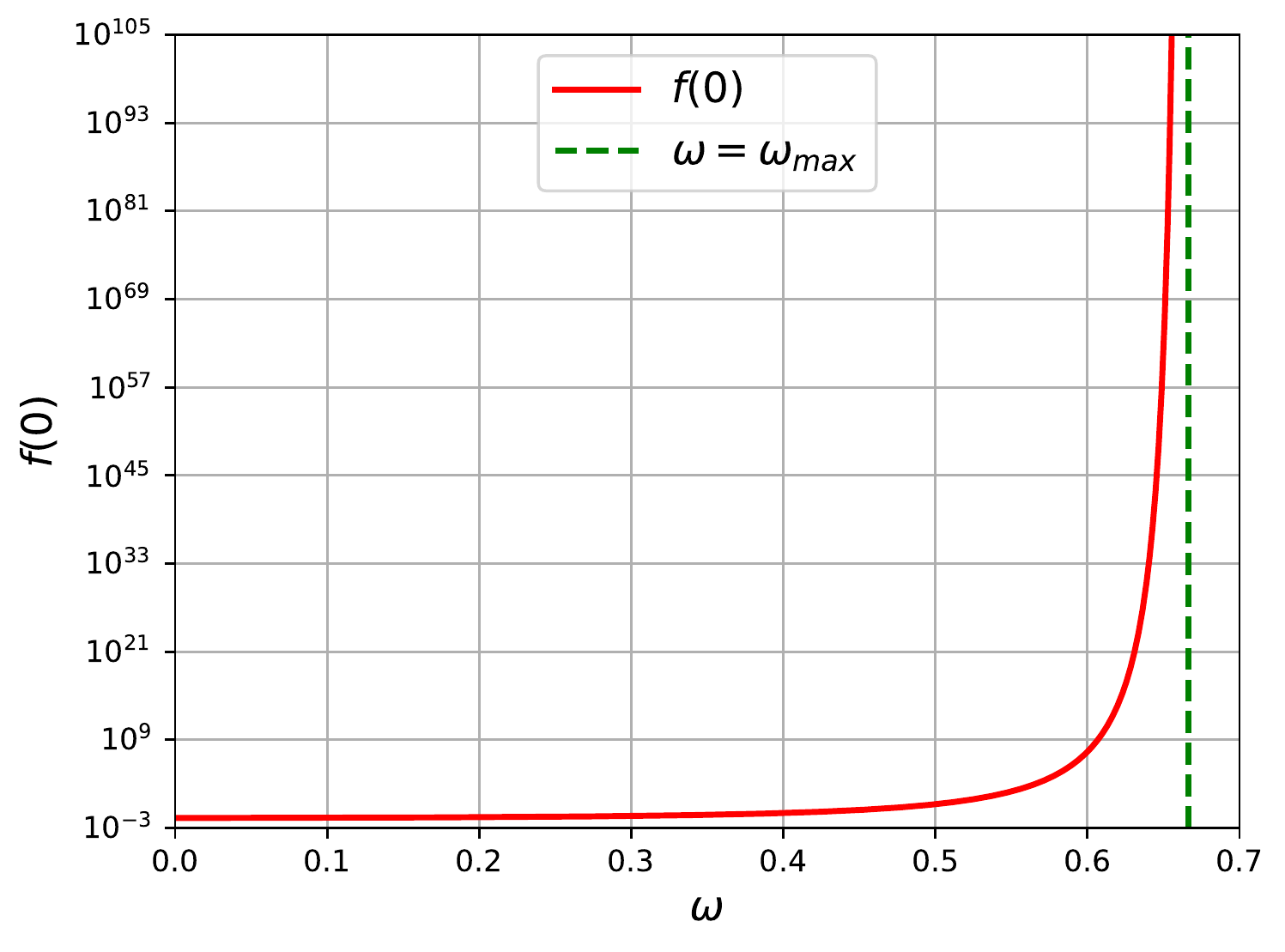}
\par\end{centering}
\caption{The values of the self-similar solution at the origin, $f\left(0\right)$
(in red), as a function of $\omega$. The problem's parameters are
the same as in Fig. \ref{fig:f_various_lin} (linear conduction).
Negative $\omega$ values, in a wide range, are plotted on the left
figure, while positive values are plotted on the right figure. The
asymptotic approximation (equations \eqref{eq:fste_nonlin}, \eqref{eq:Blin})
of $f\left(0\right)$ for $\omega\rightarrow-\infty$ is given in
the purple dashed line. The special limiting value $\omega_{\text{max}}$
(see table \ref{tab:lin_behaviour}), is shown in the vertical line.
\label{fig:f0_xsi0_lin}}
\end{figure*}

Figs \ref{fig:f_various_lin}-\ref{fig:f0_xsi0_lin} are calculated
for the following case study: $d=3$ (spherical symmetry), $n=0$,
$\lambda=1$ and $\mu=0$, for which $\omega_{\text{max}}=\omega_{c}=\frac{2}{3}$,
and $\omega_{0}=\frac{1}{3}$.

\begin{table}
\centering{}%
\begin{tabular}{|c|c|c|c|c|c|}
\hline 
$ $ & $\omega\rightarrow-\infty$ & $\omega<\omega_{0}$ & $\omega=\omega_{0}$ & $\omega_{0}<\omega<\omega_{\text{max}}$ & $\omega\rightarrow\omega_{\text{max}}^{-}$\tabularnewline
\hline 
\hline 
$f\left(0\right)$ & $\infty$ & finite & finite & finite & $\infty$\tabularnewline
\hline 
$f'\left(0\right)$ & 0 & 0 & finite & $-\infty$ & $-\infty$\tabularnewline
\hline 
\end{tabular}\caption{Behavior of the the self-similar solution $f\left(0\right)$ and its
derivative $f'\left(0\right)$ at the origin, for various ranges of
the spatial density power $\omega$, for linear conduction ($n=0$).\label{tab:lin_behaviour}}
\end{table}

\section{Comparison with simulations\label{sec:simulations}}

\begin{figure*}[t]
\begin{centering}
\includegraphics[scale=0.42]{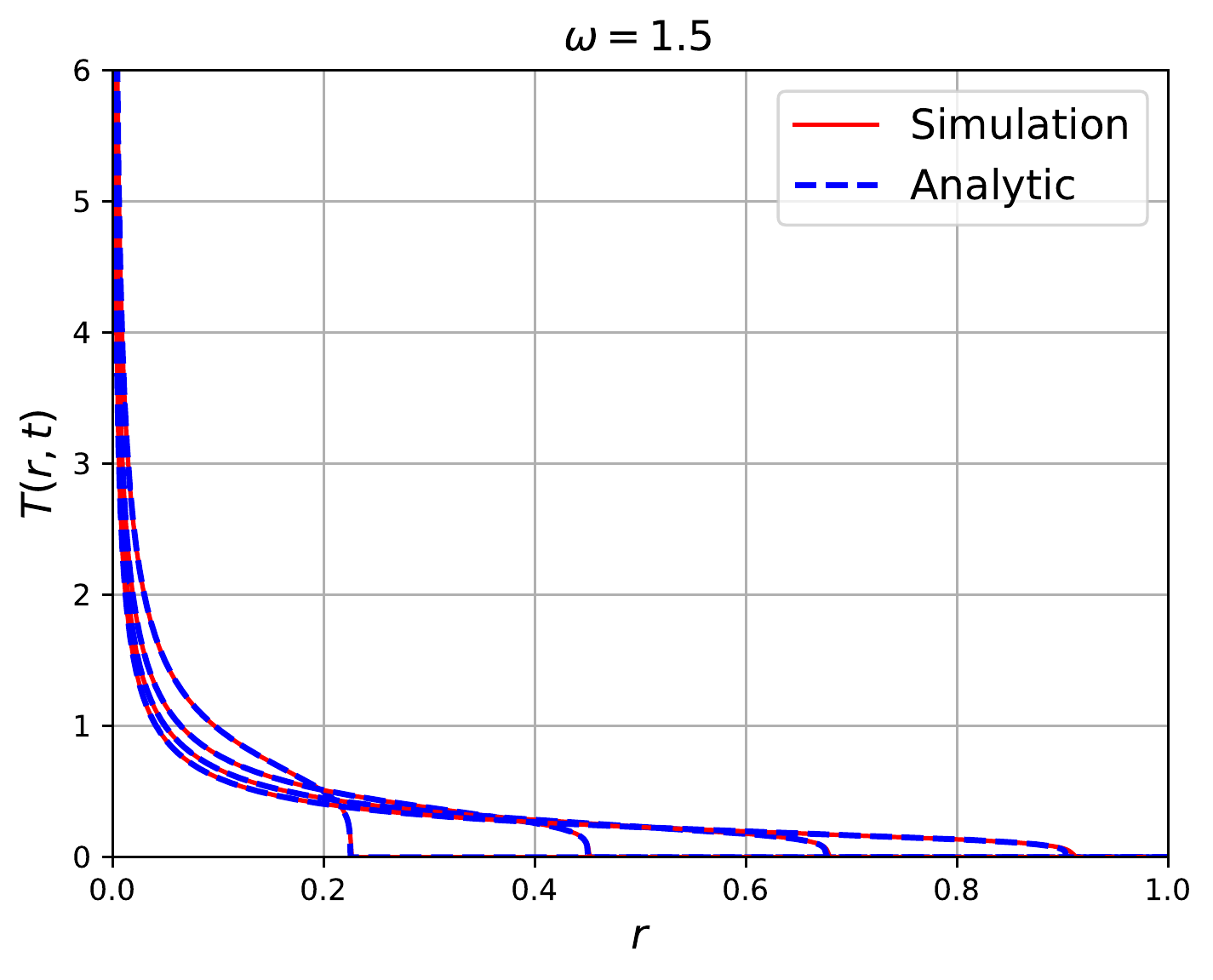}\includegraphics[scale=0.42]{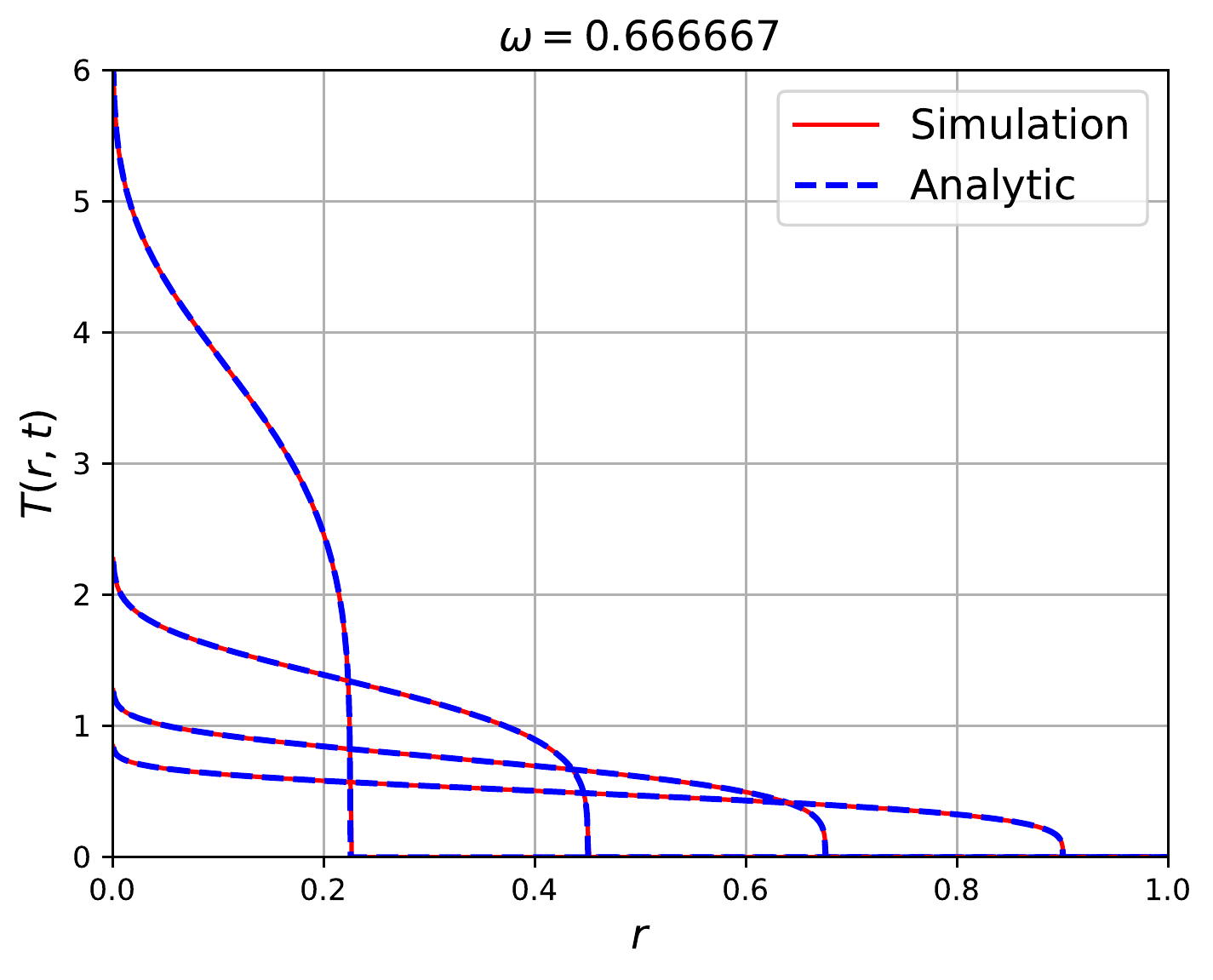}\includegraphics[scale=0.42]{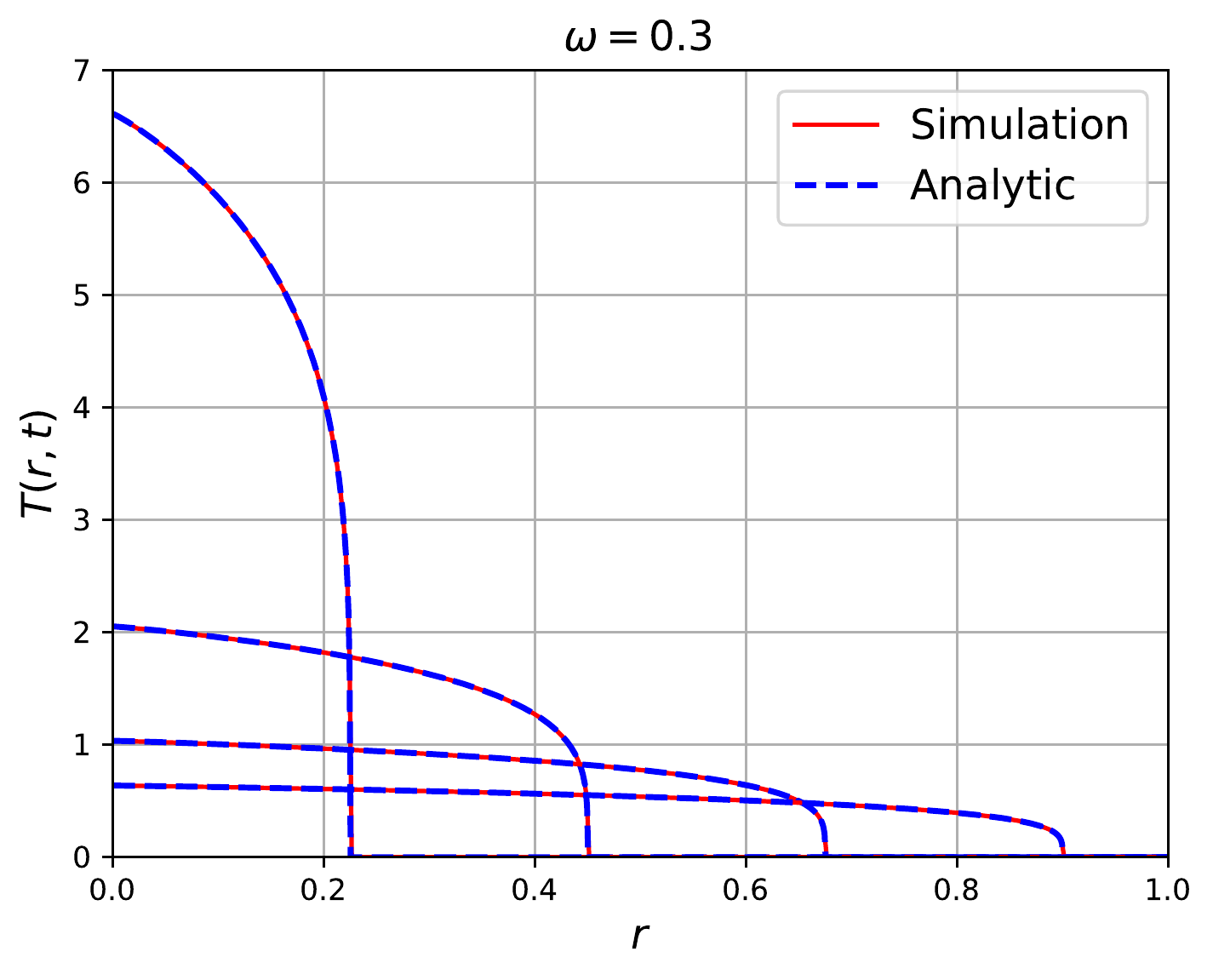}
\par\end{centering}
\begin{centering}
\includegraphics[scale=0.42]{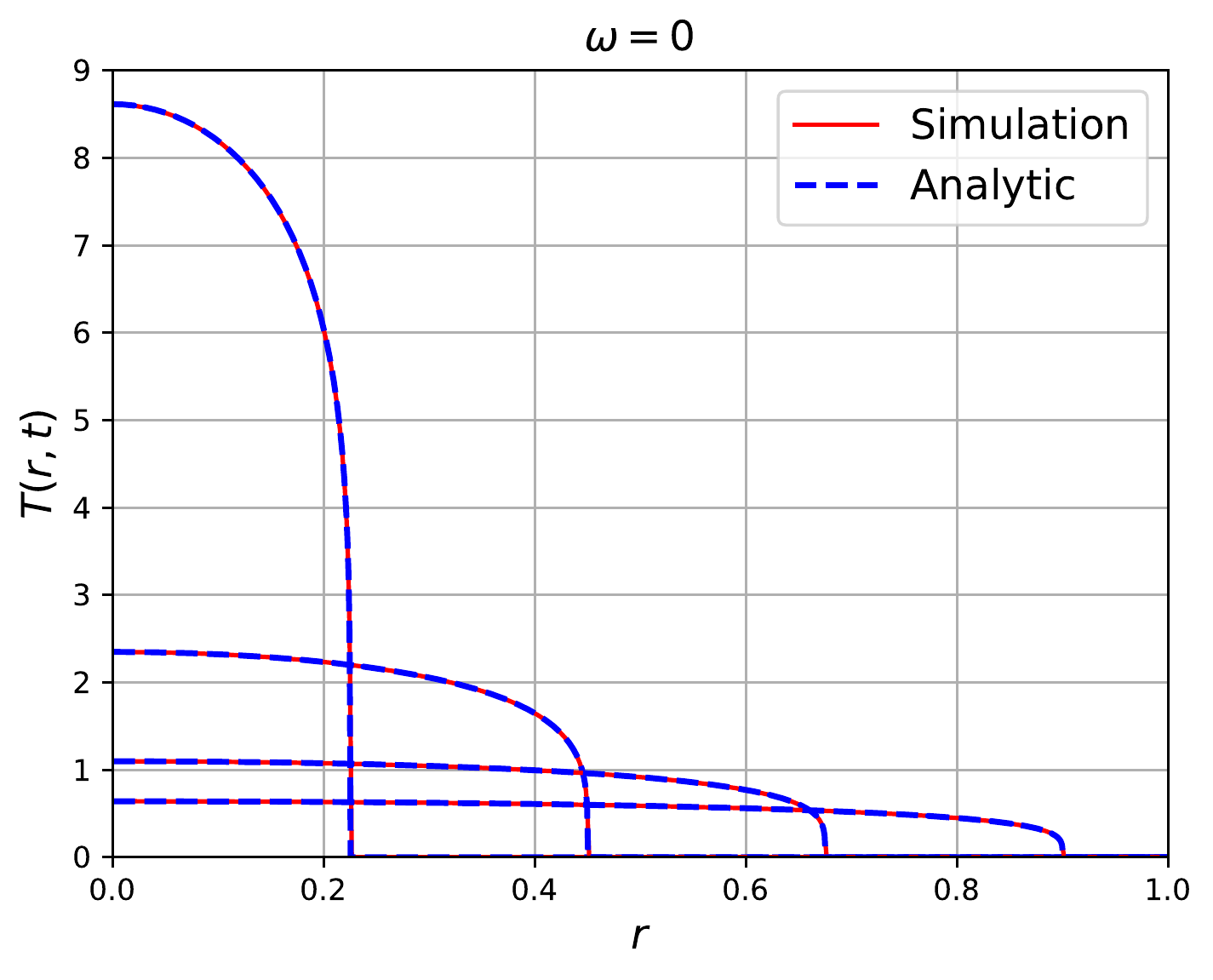}\includegraphics[scale=0.42]{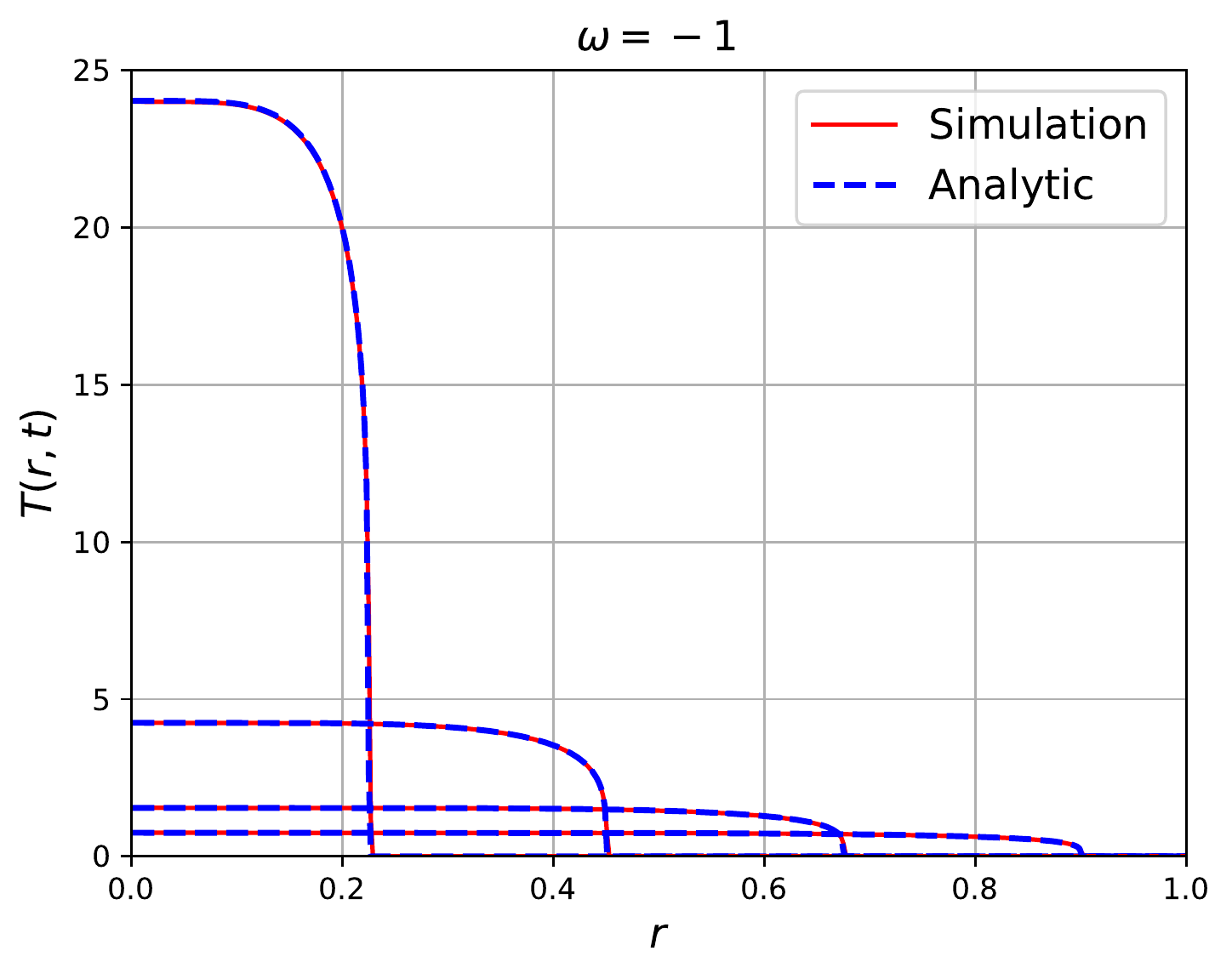}\includegraphics[scale=0.42]{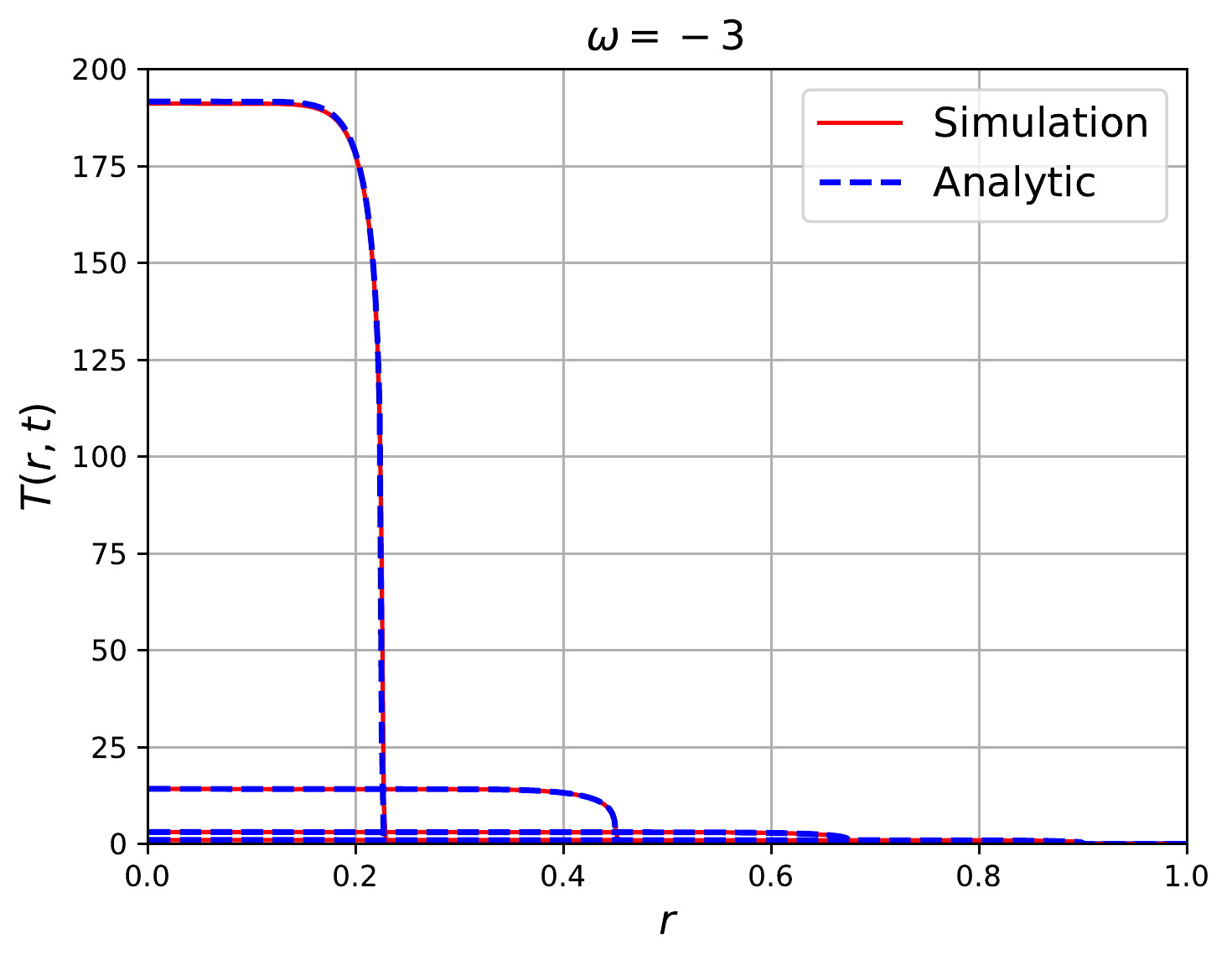}
\par\end{centering}
\caption{A comparison between numerical simulations (red lines) and analytic
solutions (blue dashed lines) of the radiation diffusion equation
\eqref{eq:main_eq} with an instantaneous point source for various
values of $\omega$ (given in the title of each sub-figure). Temperature
profiles (in Kelvin) are shown at different times such that
the heat wave reaches radii of $r=$0.225, 0.45, 0.675 and 0.9. The
comparisons are performed for $d=3$ (spherical symmetry), $\alpha=2$,
$\beta=1.6$, $\lambda=1$ and $\mu=0$, for which $n=2.75$ (nonlinear
conduction), $\omega_{\text{max}}=1.7826$, $\omega_{\text{acc}}=1.6087$,
$\omega_{c}=\frac{2}{3}$ and $\omega_{0}=\frac{1}{3}$. The energy
deposited at the origin at $t=0$ is $Q=1$. All quantities are given
in c.g.s. units. \label{fig:simulation_nonlin}}
\end{figure*}

\begin{figure*}[t]
\begin{centering}
\includegraphics[scale=0.42]{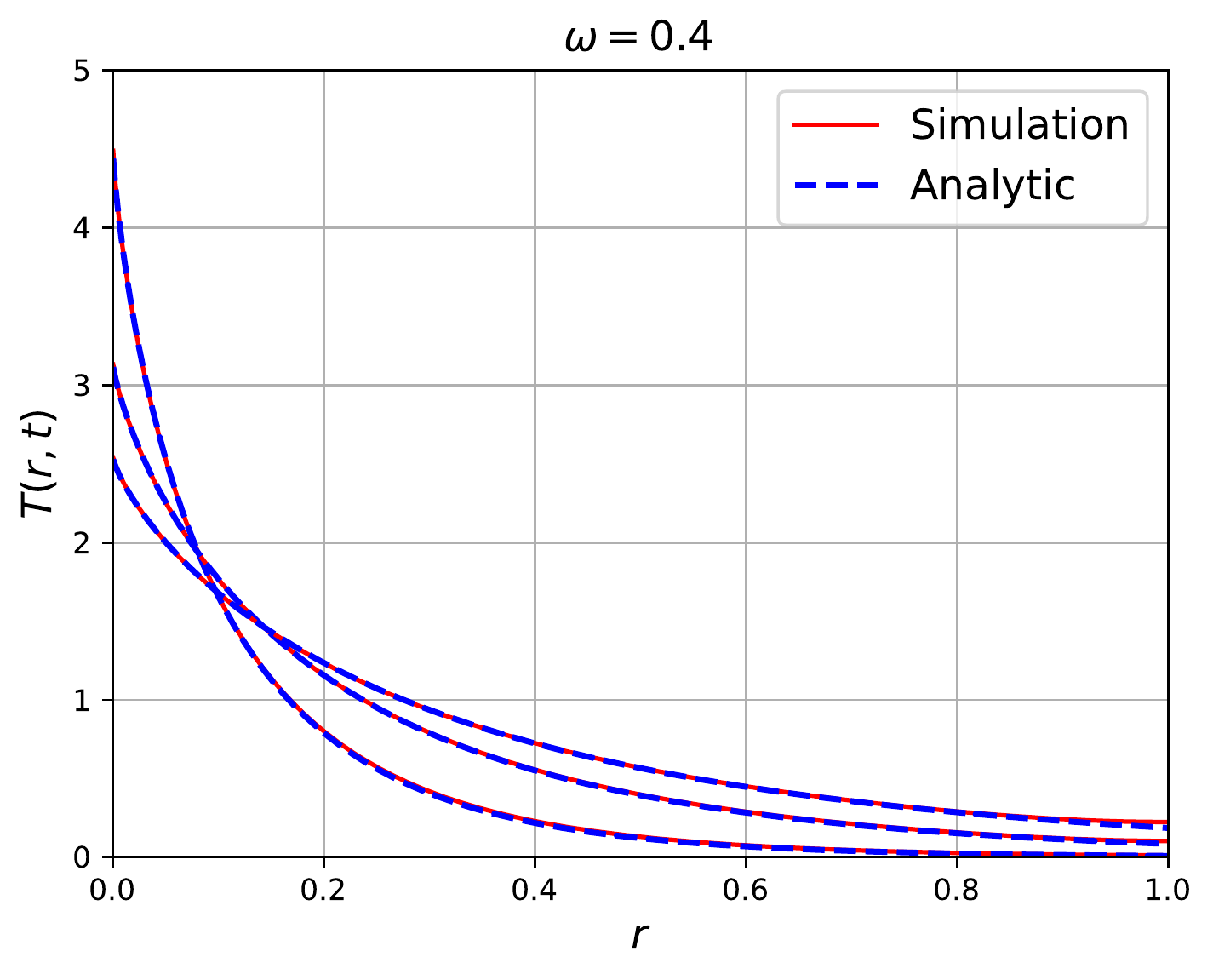}\includegraphics[scale=0.42]{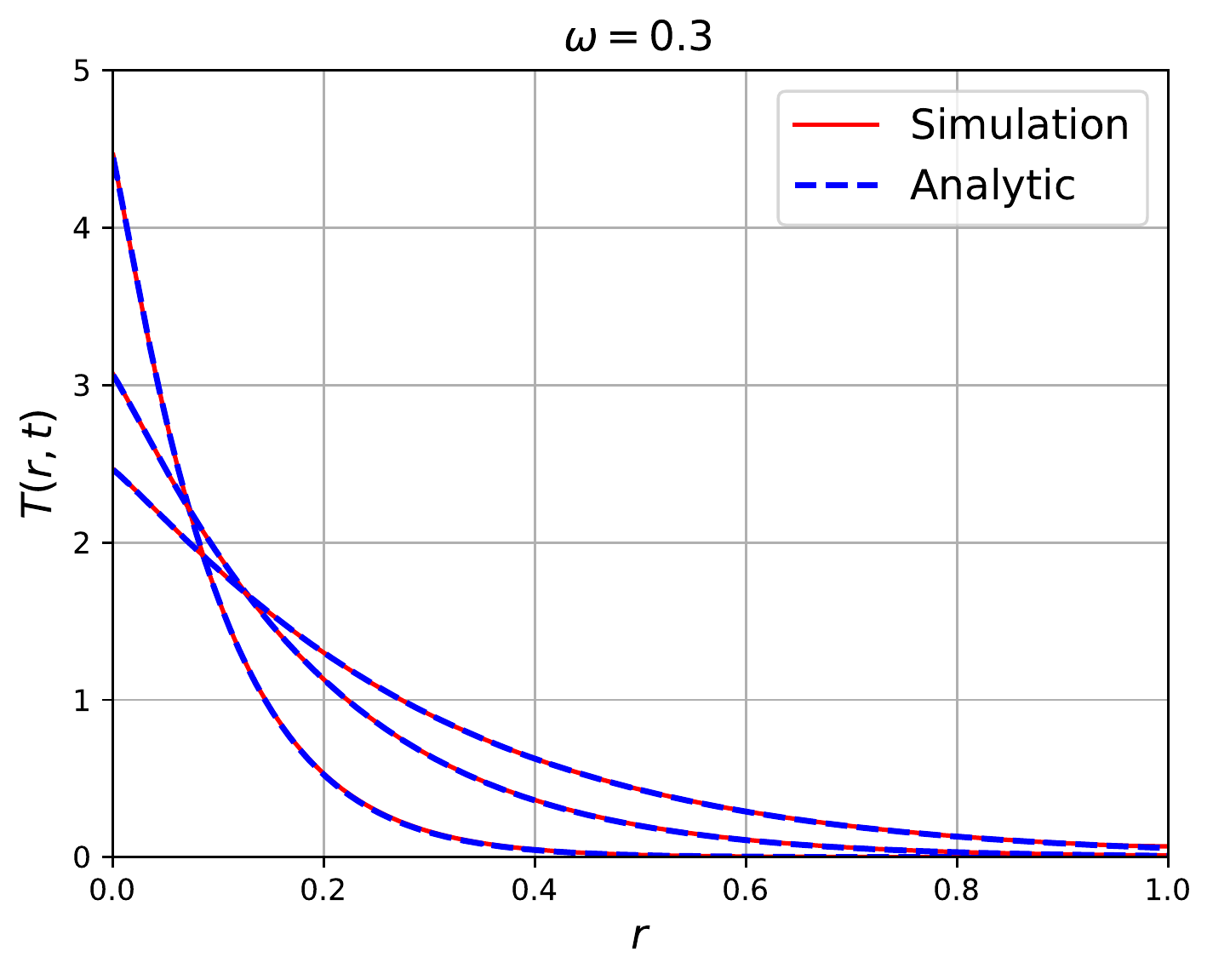}\includegraphics[scale=0.42]{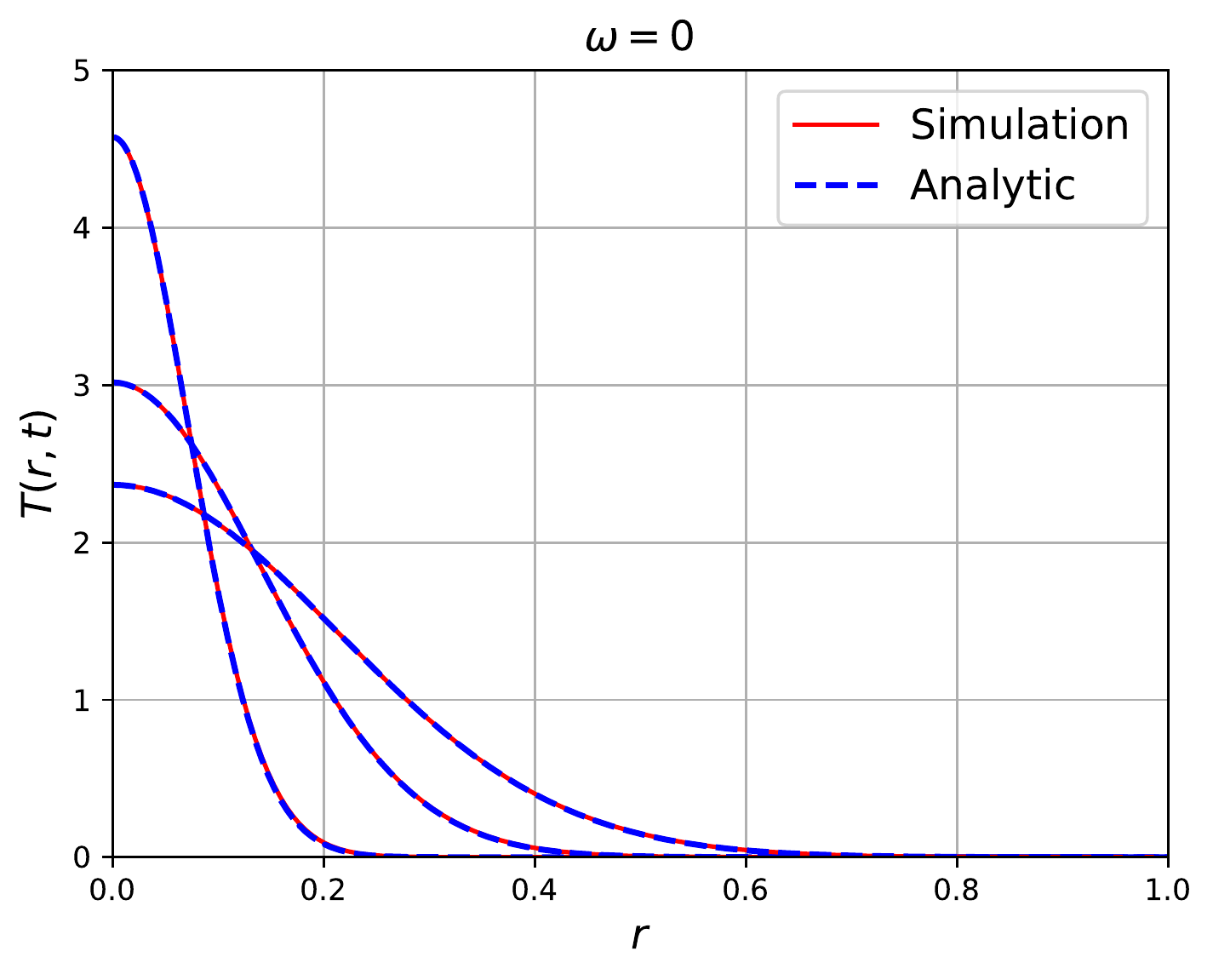}
\par\end{centering}
\begin{centering}
\includegraphics[scale=0.42]{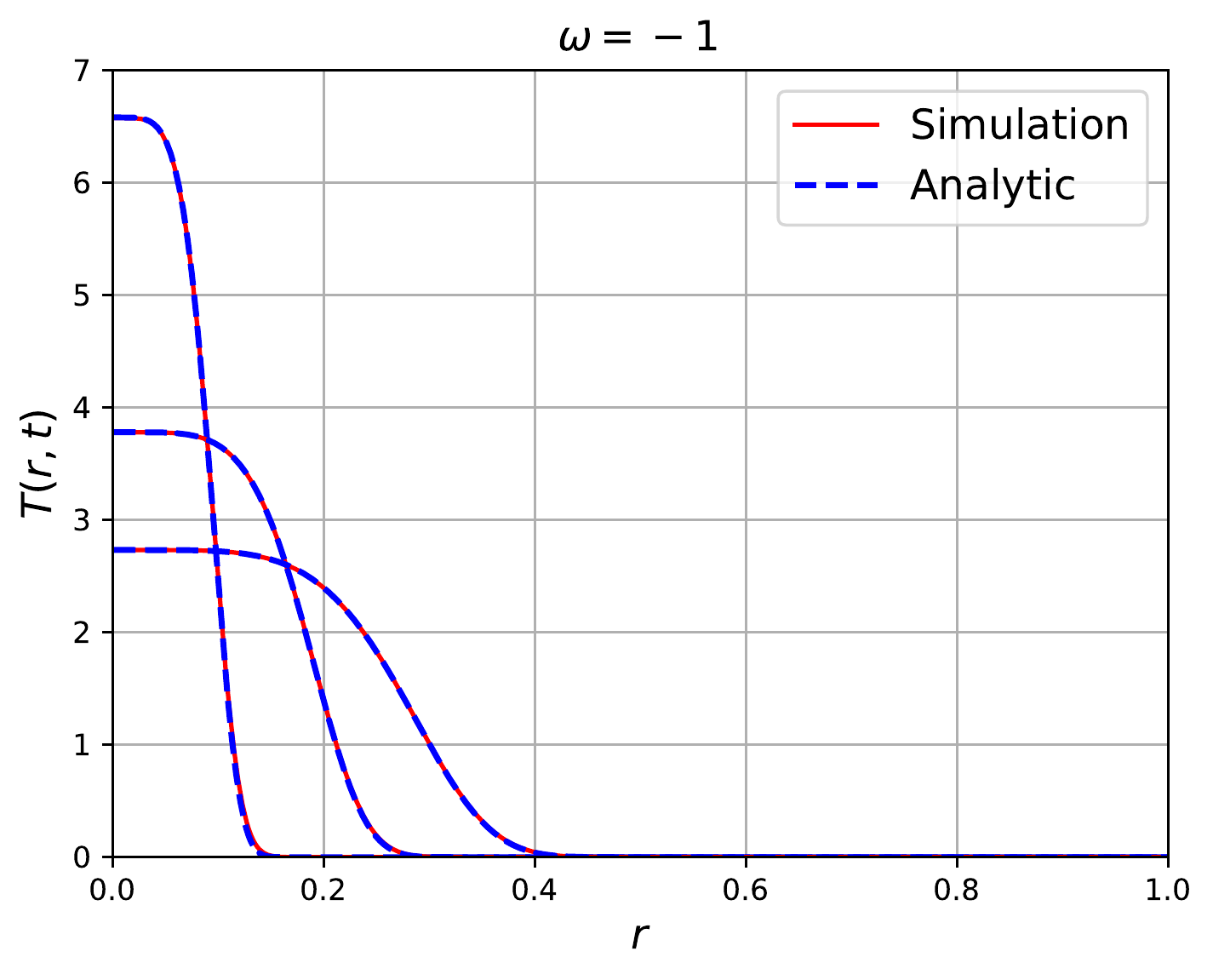}\includegraphics[scale=0.42]{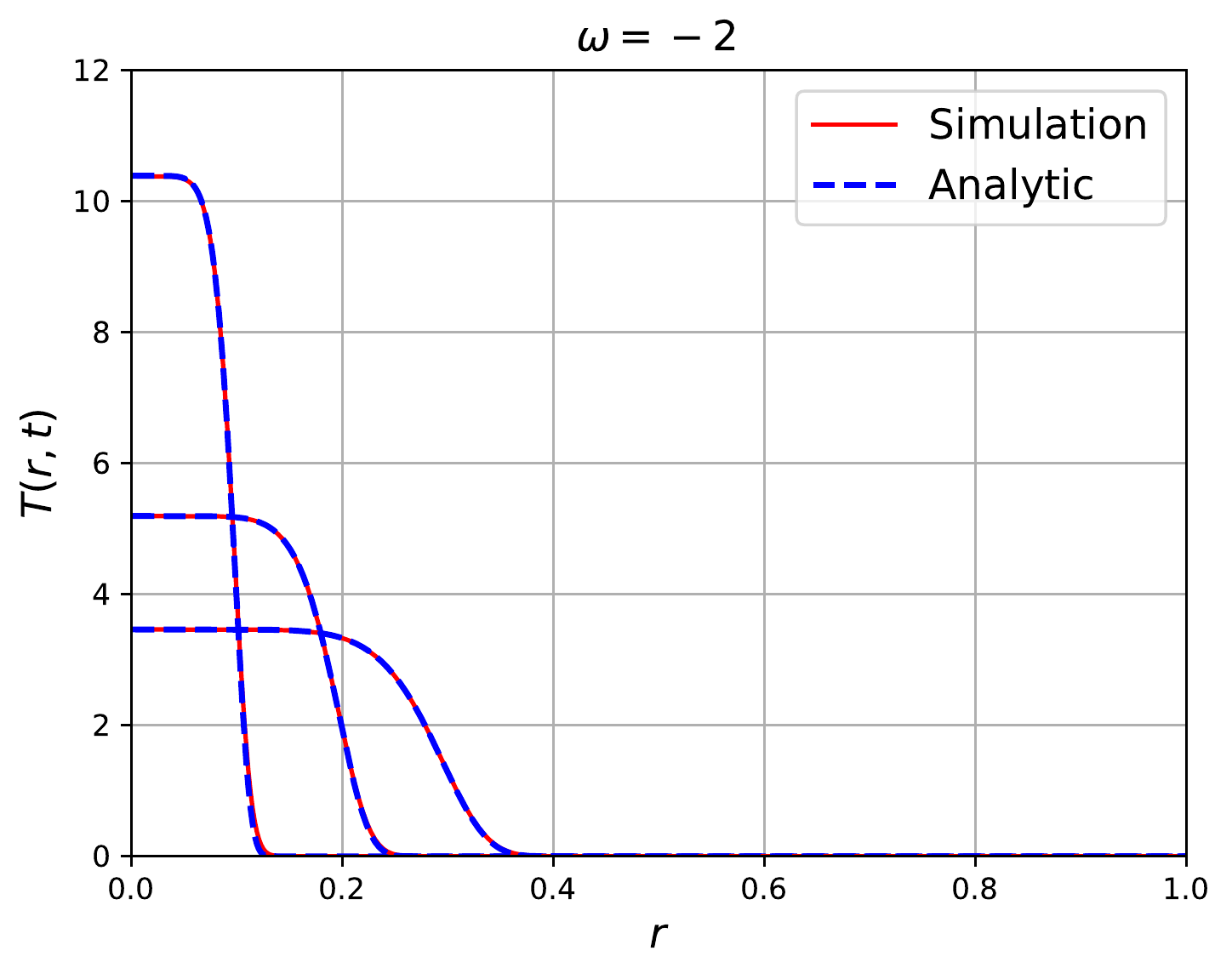}\includegraphics[scale=0.42]{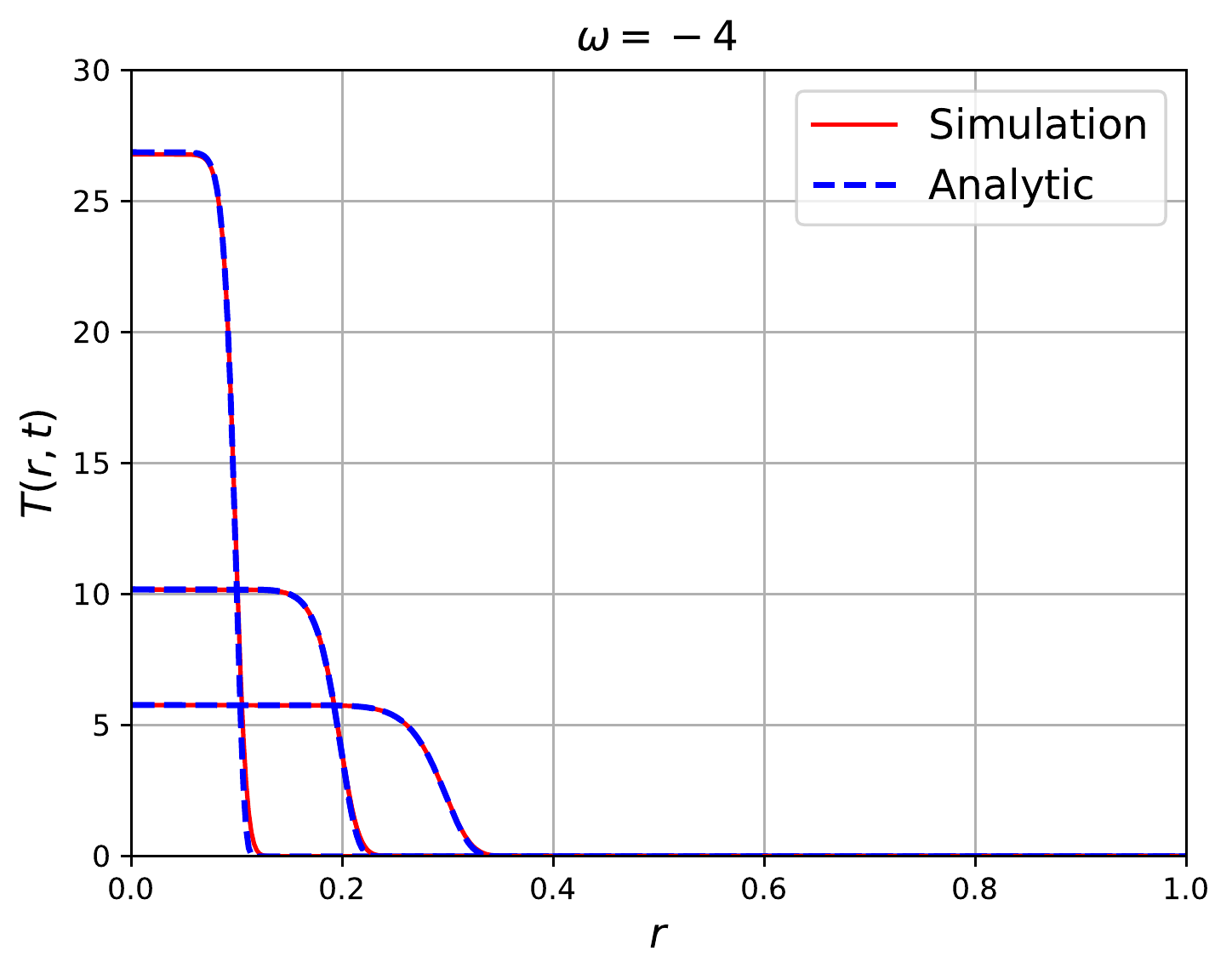}
\par\end{centering}
\caption{A comparison between numerical simulations (red lines) and analytic
solutions (blue dashed lines) of the radiation diffusion equation
\eqref{eq:main_eq} with an instantaneous point source for various
values of $\omega$ (given in the title of each sub-figure). Temperature
profiles are shown at different times such that the argument of the
exponent in Eq. \eqref{eq:fxsi_anal_linear} equals 0.1333, 0.2667
and 0.4. The comparisons are performed for $d=3$ (spherical symmetry),
$\alpha=1$ and $\beta=5$, $\lambda=1$ and $\mu=0$, for which $n=0$
(linear conduction), $\omega_{\text{max}}=\omega_{c}=\frac{2}{3}$,
and $\omega_{0}=\frac{1}{3}$. \label{fig:simulation_lin}}
\end{figure*}

\begin{figure}
\begin{centering}
\includegraphics[scale=0.5]{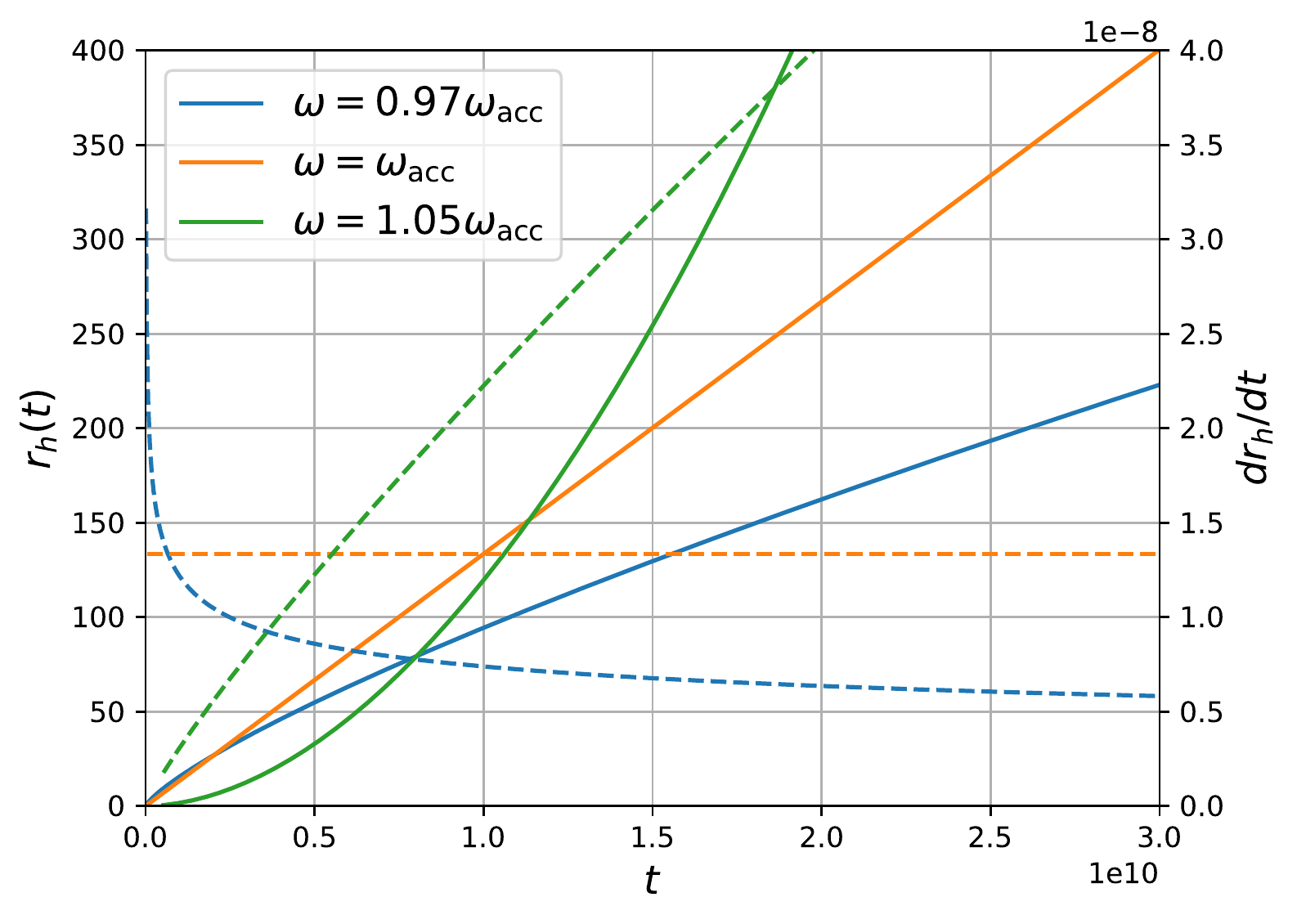}
\par\end{centering}
\caption{The heat wave position in Eq. \eqref{eq:hw_nonlin} (left y axis,
solid lines) and velocity (right y axis, dashed lines) as a function
of time, for the same parameters of Fig. \ref{fig:simulation_nonlin}
(nonlinear conduction), and for $\omega=0.97\omega_{\text{acc}}=1.56$
(decelerating heat front, in blue), $\omega=\omega_{\text{acc}}=1.609$
(constant speed heat front, in orange) and $\omega=1.05\omega_{\text{acc}}=1.689$
(accelerating heat front, in orange). \label{fig:r_hw}}
\end{figure}

In order to test the new analytic solutions presented in this work
and to demonstrate their utility for verification and validation of
computer simulations, we have performed detailed numerical simulations
of the radiation diffusion equation \eqref{eq:main_eq} for both nonlinear
and linear conduction and for various values of $\omega$. The simulations
use a standard one dimensional fully implicit nonlinear diffusion
scheme, and temperature time step control with a tolerance of $1\%$.
All simulations are performed for spherical symmetry ($d=3$), using
1000 computational cells, with power law opacity (Eq. \eqref{eq:ross_opac_powerlaw}
with $g=1$), and EOS (Eq. \eqref{eq:eos} with $f=1$), and are initialized
with a spatial power law density (Eq. \eqref{eq:rho_powerlaw} with
$\rho_{0}=1$), and with a total energy of $Q=1$ deposited in the
first cell (and zero energy elsewhere), in order to represent the
instantaneous energy point source. All quantities are given in c.g.s.
units.

In Fig. \ref{fig:simulation_nonlin}, the numerical simulations are
compared with the analytic solutions for nonlinear conduction given
in Eq. \eqref{eq:fxsi_sol} for the parameters $\alpha=2$, $\beta=1.6$,
$\lambda=1$ and $\mu=0$, for which $n=2.75$, $\omega_{\text{max}}=1.7826$,
$\omega_{\text{acc}}=1.6087$, $\overline{\omega}=1.4762$, $\omega_{c}=\frac{2}{3}$
and $\omega_{0}=\frac{1}{3}$ (which are the same parameters of the
nonlinear conduction case studied in Figs. \ref{fig:f_various}-\ref{fig:f0_xsi0}).
Temperature profiles are shown at different times such that the nonlinear
heat wave (Eq. \eqref{eq:hw_nonlin}) reaches radii of $r=0.2$, 0.5,
0.7 and 0.9. The values of $\omega$ were chosen in order to present
the different solution forms, according to table \ref{tab:nonlin_behaviour}.
A comparison for the special value $\omega=\omega_{c}$ is also shown,
which validates and demonstrates the marginal self-similar solution
in Eq. \eqref{eq:omega_c_sol}. Finally, we note that the steepening
of the heat wave towards a step function is evident for increasing
negative values of $\omega$.

Similarly, in Fig. \ref{fig:simulation_lin}, the numerical simulations
are compared with the analytic solutions for linear conduction given
in Eq. \eqref{eq:fxsi_anal_linear}, for the parameters $\alpha=1$,
$\beta=5$, $\lambda=1$ and $\mu=0$, for which $n=0$, $\omega_{\text{max}}=\omega_{c}=\frac{2}{3}$,
and $\omega_{0}=\frac{1}{3}$ (which are the same parameters of the
linear conduction case studied in Figs. \ref{fig:f_various_lin}-\ref{fig:f0_xsi0_lin}).
Temperature profiles are shown at different times such that the argument
of the exponent in Eq. \eqref{eq:fxsi_anal_linear} equals 0.1333,
0.2667 and 0.4, which represent different positions the non-sharp
linear conduction heat front. The values of $\omega$ are chosen in
order to present the different solution forms, according to table
\ref{tab:lin_behaviour}. We note that the steepening of the super-Gaussian
heat wave towards a step function is evident for increasing negative
values of $\omega$.

A comparison of the nonlinear heat front position and velocity for
$\omega<\omega_{\text{acc}}$, $\omega=\omega_{\text{acc}}$ and $\omega>\omega_{\text{acc}}$
is shown in Fig. \ref{fig:r_hw}. As was argued at the end of Sec.
\ref{sec:Self-Similar-solution}, it is evident that the resulting
heat front propagation decelerates, has a constant speed and accelerates,
respectively. 

Finally, we note that we have also compared the energy fluxes given
from the simulations with the analytic solutions, presented here in
Eq. \eqref{eq:flux_selfsim}. These comparison are not presented here
for the sake of brevity, but we note that a very good agreement was
reached, so that Eq. \eqref{eq:flux_selfsim} can also be used for
the purpose of verification and validation of computer simulations.

\section{Summary}

In this work we have generalized the well known solutions to the nonlinear
radiation diffusion equation with an instantaneous point source, to
a non-homogeneous medium with a power law spatial density profile.
It was shown that the solutions take various qualitatively different
forms according to the value of the spatial density exponent $\omega$.
The different forms were studied in detail for both linear and non
linear heat conduction. The various solution forms were compared in
detail to numerical simulations, and a good agreement was achieved.
These new solutions can be used for verification and validation of
numerical simulations of the radiation diffusion equation, as well
as for any diffusion model.
\begin{acknowledgments}
The author thanks Shay I. Heizler for useful suggestions and comments.
\end{acknowledgments}

\subsection*{Availability of data}

The data that support the findings of this study are available from
the corresponding author upon reasonable request.

\bibliographystyle{unsrt}
\bibliography{datab}

\pagebreak{}

\appendix

\section{Dimensional analysis\label{sec:Dimensional-analysis}}

\begin{table}[H]
\centering{}%
\begin{tabular}{|c|c|c|c|c|}
\hline 
$w$ & $Q$ & $A$ & $r$ & $t$\tabularnewline
\hline 
\hline 
$\left[w\right]$ & $\left[w\right]\left[\text{length}\right]^{d-m}$ & $\frac{\left[\text{length}\right]^{2-k-m}}{\left[w\right]^{n}\left[\text{time}\right]}$ & $\left[\text{length}\right]$ & $\left[\text{time}\right]$\tabularnewline
\hline 
\end{tabular}\caption{The dimensional quantities in the problem (upper line) and their dimensions
(lower line).\label{tab:The-dimensional-quantities}}
\end{table}

In this appendix we will use the method of dimensional analysis in
order to find a self-similar ansatz for the solution of the problem
defined by equations \eqref{eq:diff_eq_1d} and \eqref{eq:cons_w}.
The dimensional quantities which define the problem are listed in
table \ref{tab:The-dimensional-quantities}. It is seen that the problem
is defined by $M=5$ dimensional quantities which are composed of
$N=3$ different units. Therefore from the central theorem of dimensional
analysis (a.k.a. the Pi theorem) \cite{buckingham1914physically,zeldovich1967physics,barenblatt1996scaling},
the problem can be solved using $M-N=2$ dimensionless variables,
written in terms of power laws of the dimensional quantities:
\begin{equation}
\xi=rQ^{x}A^{y}t^{z},\label{eq:xsiapp}
\end{equation}
\begin{equation}
f\left(\xi\right)=\frac{w}{Q^{a}A^{b}t^{c}}.\label{eq:fapp}
\end{equation}
The requirement that $\xi$ is dimensionless gives:

\begin{align*}
 & x-yn=0\\
 & 1+\left(d-m\right)x+\left(2-k-m\right)y=0\\
 & -y+z=0
\end{align*}
which has the solution:
\[
x=-\frac{n}{p},
\]

\[
y=z=-\frac{1}{p},
\]
where $p$ is given by Eq. \eqref{eq:pdef}. Similarly, the requirement
that $f\left(\xi\right)$ is dimensionless gives:

\begin{align*}
 & 1-a+bn=0\\
 & -a\left(d-m\right)-\left(2-k-m\right)b=0\\
 & b-c=0
\end{align*}
which has the solution:

\[
a=\frac{2-k-m}{p},
\]
\[
b=c=-\frac{d-m}{p}.
\]
Hence, it is seen that the resulting dimensionless quantities \eqref{eq:xsiapp}-\eqref{eq:fapp}
give the self-similar ansatz \eqref{xsi_def}-\eqref{eq:w_f}.

\section{The calculation of $\xi_{0}$\label{app:xsi0}}

The value of $\xi_{0}$ can be found by substituting Eq. \eqref{eq:fxsi_sol}
into the energy conservation constraint \eqref{eq:cons_selfsim}.
This substitution results in the following equation:

\begin{equation}
\frac{n^{\frac{1}{n}}\mathcal{A}_{d}\xi_{0}^{\frac{2-k-m}{n}+d-m}}{\left(p\left|2-k-m\right|\right)^{\frac{1}{n}}}I=1,\label{eq:xsi0_eq}
\end{equation}
where $I$ is the following definite integral:
\begin{equation}
I=\int_{0}^{1}x^{d-m-1}\left|1-x^{2-k-m}\right|^{\frac{1}{n}}dx.\label{eq:Idef}
\end{equation}
Under the conditions $2-k-m>0$, $d-m>0$ and $\frac{1}{n}+1>0$,
this integral is well known (i.e. Ref. \cite{gradshteyn2014table},
page 324, Eq. 3.251-1), and given by:
\begin{align}
I & =\frac{1}{2-k-m}\mathcal{B}\left(\frac{d-m}{2-k-m},\frac{1}{n}+1\right),\label{eq:Ipos}
\end{align}
with the Beta function defined in terms of the Gamma function:
\[
\mathcal{B}\left(x,y\right)=\frac{\Gamma\left(x\right)\Gamma\left(y\right)}{\Gamma\left(x+y\right)}.
\]
On the other hand, if  $2-k-m<0$, we employ the integral
(see i.e. Ref. \cite{gradshteyn2014table} page 325, Eq. 3.251-3):
\[
\int_{1}^{\infty}x^{s-1}\left(x^{q}-1\right)^{v-1}dx=\frac{1}{q}\mathcal{B}\left(1-v-\frac{s}{q},v\right),
\]
which is valid for $q>0$, $v>0$, and $s<q\left(1-v\right)$, so
that after a change of variables $y=\frac{1}{x}$ in Eq. \eqref{eq:Idef},
one finds that:

\begin{equation}
I=-\frac{1}{2-k-m}\mathcal{B}\left(-\frac{1}{n}-\frac{d-m}{2-k-m},\frac{1}{n}+1\right),\label{eq:Ineg}
\end{equation}
which is valid under the condition $\frac{1}{n}+1>0$ and $2-k-m+\left(d-m\right)n>0,$which
is already assumed in Eq. \eqref{eq:2km_constraint}. Using the results
in equations \eqref{eq:Ipos},\eqref{eq:Ineg} in Eq. \eqref{eq:xsi0_eq}
and solving for $\xi_{0}$, results in equations \eqref{eq:xsi0_final}-\eqref{eq:l_xsi_0}.

\section{Analytic solution for $\omega=\omega_{c}$\label{sec:Analytic-solution-foroemgac}}

For $\omega=\omega_{c}$ (for which $2-k-m=0$), the ODE \eqref{eq:ode}
reads:
\[
nf^{n-1}\left(\xi\right)f'\left(\xi\right)=-\frac{1}{\left(d-m\right)\xi}.
\]
Assuming $n>0$, and employing the boundary condition at infinity
(see Eq. \eqref{eq:fxsi_inf}), gives the solution:
\[
f\left(\xi\right)=\begin{cases}
\left[\frac{\ln\left(\frac{\xi_{0}}{\xi}\right)}{d-m}\right]^{\frac{1}{n}}, & \xi<\xi_{0}\\
0, & \xi>\xi_{0}
\end{cases}
\]
where $\xi_{0}$ is a constant of integration, which can be obtained,
as done in Appendix \eqref{app:xsi0}, by employing the energy conservation
constraint \eqref{eq:cons_selfsim}. This gives:
\begin{equation}
\frac{\mathcal{A}_{d}\xi_{0}^{d-m}}{\left(d-m\right)^{\frac{1}{n}}}I=1,\label{eq:app2}
\end{equation}
where $I$ is the following definite integral:
\begin{align}
I & =\int_{0}^{1}\left[\ln\left(\frac{1}{x}\right)\right]^{\frac{1}{n}}x^{d-m-1}dx=\frac{\Gamma\left(1+\frac{1}{n}\right)}{\left(d-m\right)^{1+\frac{1}{n}}}.\label{eq:Iapp}
\end{align}
where we have used a well known integral identity (i.e. Ref. \cite{gradshteyn2014table}
page 551, Eq. 4.272-6), which is valid under the condition $d-m>0$
(which must hold for $\omega=\omega_{c}$, since we assume Eq. \eqref{eq:2km_constraint}).
Using Eq. \eqref{eq:Iapp} in Eq. \eqref{eq:app2} and solving for
$\xi_{0}$, results in Eq. \eqref{eq:xsi0_omegac}.
\end{document}